\documentclass[%
 reprint,
superscriptaddress,
 amsmath,amssymb,
 aps,
floatfix,
]{revtex4-2}

\usepackage{graphicx}
\usepackage{dcolumn}
\usepackage{bm}
\usepackage{times}
\usepackage[usenames,dvipsnames]{xcolor}
\usepackage[colorlinks=true, linkcolor=Blue, citecolor=Blue, urlcolor=Blue]{hyperref} 
\usepackage{scalerel}
\usepackage{stackengine}

\newcommand\reallywidetilde[1]{\ThisStyle{%
  \setbox0=\hbox{$\SavedStyle#1$}%
  \stackengine{-.1\LMpt}{$\SavedStyle#1$}{%
    \stretchto{\scaleto{\SavedStyle\mkern.2mu\sim}{.5467\wd0}}{.7\ht0}%
  }{O}{c}{F}{T}{S}%
}}

\usepackage{color}
\newcommand{\Add}[1]{\textcolor{black}{#1}}

\begin{document}

\preprint{APS/123-QED}

\title{Realization of High-Fidelity CZ Gate based on a Double-Transmon Coupler}


\author{Rui Li}
\email{rui.li.gj@riken.jp}
\affiliation{RIKEN Center for Quantum Computing (RQC), Wako, Saitama 351-0198, Japan}
\author{Kentaro Kubo}
\affiliation{Frontier Research Laboratory, Corporate Research \& Development Center,Toshiba Corporation, Saiwai-ku, Kawasaki 212-8582, Japan}
\author{Yinghao Ho}
\affiliation{Frontier Research Laboratory, Corporate Research \& Development Center,Toshiba Corporation, Saiwai-ku, Kawasaki 212-8582, Japan}
\author{Zhiguang Yan}
\affiliation{RIKEN Center for Quantum Computing (RQC), Wako, Saitama 351-0198, Japan}
\author{Yasunobu Nakamura}
\email{yasunobu@ap.t.u-tokyo.ac.jp}
\affiliation{RIKEN Center for Quantum Computing (RQC), Wako, Saitama 351-0198, Japan}
\affiliation{Department of Applied Physics, Graduate School of Engineering, The University of Tokyo, Bunkyo-ku, Tokyo 113-8656, Japan}
\author{Hayato Goto}
\email{hayato1.goto@toshiba.co.jp}
\affiliation{Frontier Research Laboratory, Corporate Research \& Development Center,Toshiba Corporation, Saiwai-ku, Kawasaki 212-8582, Japan}
\affiliation{RIKEN Center for Quantum Computing (RQC), Wako, Saitama 351-0198, Japan}




\date{\today}

\begin{abstract}
Striving for higher gate fidelity is crucial not only for enhancing existing noisy intermediate-scale quantum~(NISQ) devices but also for unleashing the potential of fault-tolerant quantum computation through quantum error correction. A recently proposed theoretical scheme, the double-transmon coupler (DTC), aims to achieve both suppressed residual interaction and a fast high-fidelity two-qubit gate simultaneously, particularly for highly detuned qubits. Harnessing the state-of-the-art fabrication techniques and a model-free pulse-optimization process based on reinforcement learning, we translate the theoretical DTC scheme into reality, attaining fidelities of \Add{99.90\%} for a CZ gate and 99.98\% for single-qubit gates. The performance of the DTC scheme demonstrates its potential as a competitive building block for superconducting quantum processors.
\end{abstract}

\maketitle


\section{Introduction}\label{Introduction}

Navigating the era of Noisy Intermediate-Scale Quantum~(NISQ) devices~\cite{Preskill2018quantumcomputingin} and pioneering the era of fault-tolerant quantum computing~\cite{shor1996fault}, the blooming of qubit technologies, including superconducting qubits~\cite{nakamura1999coherent, arute2019quantum, kim2023evidence}, trapped ions~\cite{PhysRevLett.74.4091, benhelm2008towards}, neutral atoms~\cite{browaeys2016experimental, bluvstein2024logical}, photonics~\cite{zhong2020quantum} and other solid-state qubits~\cite{loss1998quantum, hendrickx2021four, rong2015experimental}, have made substantial strides toward the exciting goal of realizing quantum advantage. Despite differing entities for qubit realization, achieving high-fidelity two-qubit gates stands as a pivotal and shared challenge. While considerable efforts have been invested in enhancing gate performance~\cite{gaebler2016high, PhysRevLett.117.060504, 2023Natur.622..268E, xue2022quantum, noiri2022fast, xie202399}, superconducting qubits~\cite{PhysRevX.13.031035, PhysRevLett.127.080505, PhysRevX.11.021058, PhysRevLett.126.220502, PRXQuantum.4.010314}, in particular, exhibits promising potential due to its scalability~\cite{arute2019quantum, krinner2019engineering} and modularity~\cite{gold2021entanglement, niu2023low}.


Within the realm of superconducting qubits, the transmon~\cite{koch2007charge} has become a favorite in both academia and industry. This preference is rooted in its high coherence, achieved through cutting-edge design~\cite{koch2007charge, wang2015surface, martinis2022surface} and fabrication~\cite{place2021new, wang2022towards}, coupled with its inherent simplicity that facilitates easy controllability~\cite{schreier2008suppressing}. Diverse schemes based on transmons for implementing two-qubit gates have been proposed, encompassing fixed-frequency qubits~\cite{rigetti2010fully} and tunable alternatives~\cite{dicarlo2009demonstration}, each presenting a unique array of advantages and challenges. Fixed coupling with fixed-frequency qubits, whether through direct~\cite{patterson2019calibration} or indirect~\cite{chow2011simple} capacitive coupling, provides the feasibility of a straightforward microwave-driven two-qubit gate~\cite{magesan2020effective}. The delicate balance between high control speed and suppression of the so-called residual \textit{ZZ} interaction is carefully considered to mitigate potential gate errors~\cite{sheldon2016procedure, kandala2021demonstration}. Remarkably, tunable-coupling schemes involving a single frequency-tunable transmon coupler (STC)~\cite{yan2018tunable} have garnered attention for their ability to achieve a high on--off ratio~\cite{PhysRevLett.125.240503} of the \textit{ZZ} interactions between qubits, a critical metric for estimating the potential to simultaneously achieve high fidelity in both single- and two-qubit gates. Notably, however, the absence of the residual \textit{ZZ} interaction is confined to the so-called straddling regime~\cite{PhysRevApplied.12.054023, PhysRevApplied.16.024037}, where a detuning between qubit frequencies is smaller than their anharmonicities, potentially leading to issues with extra frequency collisions in larger systems~\cite{brink2018device, hertzberg2021laser, osman2023mitigation}.

A novel coupling scheme, employing a coupler consisting of double transmons with a shared superconducting loop and an additional Josephson junction, has been proposed theoretically to overcome the drawback above and retain the benefits of its tunability~\cite{PhysRevApplied.18.034038, kubo2023fast, campbell2023modular}. The implementation of this double-transmon-coupler~(DTC) scheme involves four transmons: two transmons function as data qubits, while the other two transmons serve as the coupler elements. The \textit{ZZ} interaction between the two data qubits can be controlled by the magnetic flux through the DTC's loop and suppressed completely even for two qubits placed outside the straddling regime without sacrificing the speed of the CZ gate \cite{PhysRevApplied.18.034038}. Moreover, the absence of the necessity for direct capacitive coupling between the two qubits positions them into a regime marked by minimal spectator error and enables flexible qubit--qubit distances, ultimately reducing both quantum and classical crosstalk~\cite{kubo2024highperformance, PhysRevApplied.12.064022, zhao2023mitigation}. 

Despite the advantages offered by the DTC scheme, achieving a high-fidelity CZ gate experimentally remains challenging, even though it has been theoretically proposed~\cite{PhysRevApplied.18.034038}. A numerically optimized pulse shape necessitates calculations based on the total energy levels of the four transmons (two data qubits and two coupler transmons). However, in practice, characterizing the full energy spectra as a function of the flux bias is a non-trivial task. This difficulty arises because strongly hybridized states between the qubits and coupler transmons at certain flux bias points hinder efficient readout. Regardless of the theoretically calculated pulse shape, opting for a commonly used Slepian pulse shape~\cite{press2007numerical, PhysRevA.90.022307, PhysRevX.11.021058} can be an alternative choice. This method does not require detailed energy-level information, although it may not necessarily be the optimal choice. However, another challenge arises from the constrained bandwidth and unwanted dispersion of the circuits for pulse transmission~\cite{rol2020time}. This distorts the ideal pulse shape applied to the DTC loop, thereby impeding the realization of the ideal adiabatic process. Although a pulse-transient calibration method has been proposed to mitigate pulse distortion~\cite{rol2020time, PhysRevX.11.021058}, reading out the states of coupler transmons require additional resonators, further burdening an already complicated and crowded on-chip circuit.

In this work, we overcome these problems and experimentally demonstrate a high-fidelity CZ gate on a two-qubit device with a DTC. We design the chip considering both high coherence and fabrication feasibility. We achieve a high on--off ratio $>$$10^4$ for the \textit{ZZ} interaction between the two qubits, a crucial factor for attaining high gate fidelities for both single- and two-qubit gates. At the idle bias point, a minimum residual \textit{ZZ} interaction~[$2\pi \times (-6.3)$~kHz] persists without compromising the single-qubit gate fidelities, which are measured to be over 99.98\% through simultaneous randomized benchmarking~\cite{PhysRevLett.109.240504}. We calibrate the distortion of the \textit{Z}-pulse applied on the DTC by utilizing the readout resonator of a qubit. Subsequently, through further optimization driven by a model-free reinforcement learning (RL) algorithm, we achieve a fast and optimized pulse for a high-fidelity CZ gate. The CZ-gate fidelity, \Add{$99.90 \pm 0.01$\%}, demonstrated through Clifford interleaved randomized benchmarking~\cite{PhysRevA.87.030301, PhysRevLett.109.080505, barends2014superconducting}, remains stable within a 12-hour timeframe. For the error budget of the CZ gate, we highlight leakage and incoherent errors as the primary contributors.

\section{Device setup}\label{Device setup} 

\begin{figure}
\includegraphics[scale=0.985]{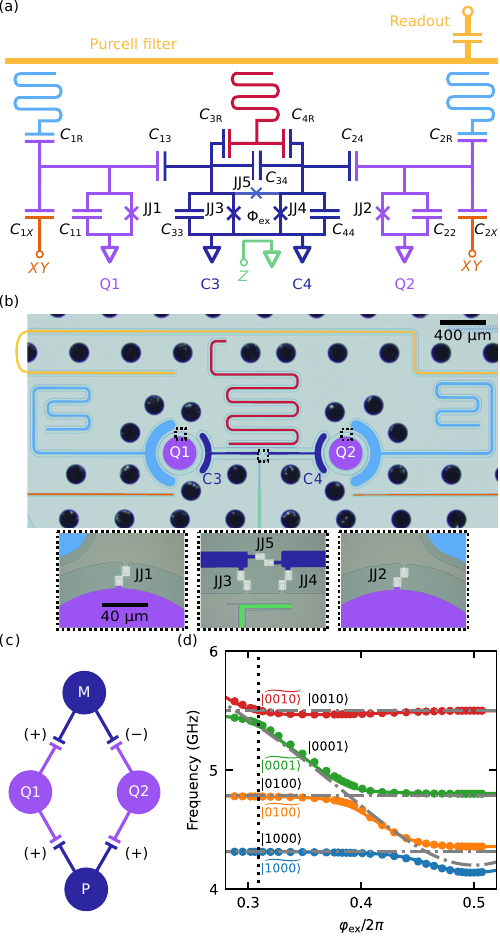}
\caption{\label{Fig1:wide} (a) Schematic circuit diagram of the DTC scheme consisting of two data qubit transmons, Q1 and Q2, and two coupler transmons, C3 and C4, as well as their readout and control elements. (b) False-color picture of the real device. The colors correspond to the circuit elements in~(a). The black holes are superconducting through-silicon vias (TSVs) distributed throughout the entire chip. The three panels at the bottom are magnified pictures of the areas~(dotted rectangles) containing Josephson junctions. (c) Conceptual sketch of a toy model of the DTC scheme. Each of Q1 and Q2 couples to the plus~(p-) and minus~(m-) modes, represented by P and M, respectively. ($+$)~and~($-$) indicate the signs of the effective couplings. (d) Energy spectra of the lowest four excited states as a function of the reduced flux bias $\varphi_\mathrm{ex}$. The dots and solid lines are experimental and simulation results, respectively. The gray dashed-dotted lines serve as guides for the four decoupled modes. The labels, $|\mathrm{Q1,}\mathrm{Q2,}\mathrm{P,}\mathrm{M}\rangle$, with~(without) a tilde are the indication of eigenstates with~(without) couplings among the four modes. The vertical dotted line indicates the idle bias point providing the minimum \textit{ZZ} interaction between Q1 and Q2.} 
\end{figure}

The demonstration device for the DTC scheme comprises four transmons: two data transmons (Q1 and Q2) as qubits and two coupler transmons (C3 and C4). \Add{The frequency detuning between Q1 (4.314 GHz) and Q2 (4.778 GHz) is 464 MHz, which exceeds the respective anharmonicities of $-212$~MHz and $-199$~MHz for Q1 and Q2 (see Appendix~\ref{Device parameters}). This configuration places the system outside the straddling regime.} Each of these transmons consists of a single Josephson junction and a large shunt capacitance. They are positioned adjacent to each other at the center of the chip [Figs.~\ref{Fig1:wide}(a) and~\ref{Fig1:wide}(b)]. Q1~(Q2) is directly coupled to C3~(C4) capacitively. While it is inevitable to have other marginal capacitive couplings between the transmons, such as the one between Q1 and Q2, they are not essential and do not significantly hinder the achievement of the designed \textit{ZZ} interaction for the CZ gate. Furthermore, a superconducting loop is formed by the Josephson junctions JJ3 and JJ4 of C3 and C4, respectively, connected by an additional Josephson junction JJ5. The DTC, represented by C3 and C4, is employed to mediate the coupling between Q1 and Q2 tuned by the magnetic flux $\Phi_\mathrm{ex}$ through the superconducting loop. 

Each qubit has its own resonator for readout, while the two coupler transmons share a third resonator. All three resonators are $\lambda/4$ coplanar waveguides, either capacitively or inductively coupled to a shared Purcell filter~\cite{houck2008controlling}. We distinguish the ground~($|0\rangle$), first excited~($|1\rangle$) and second excited states~($|2\rangle$) for each of the two qubits with the single-shot dispersive readout through their respective readout resonators. For coupler transmons, we can only discriminate the ground state $|00\rangle$ from other excited states by the readout through the shared resonator. (see Appendix~\ref{State discrimination} for details). It is important to note that the resonator for the state readout of the coupler transmons is not necessary for calibrating the high-fidelity CZ gate. However, it proves useful and sufficient for characterizing the leakage error in this work.

Each of Q1 and Q2 has its dedicated microwave control line for driving the single-qubit $X$- and $Y$-rotations. A fast flux line shorted to the ground is inductively coupled to the superconducting loop of the DTC for delivering the fast CZ-gate pulse. Additionally, in the spectroscopic measurement for characterization, we utilize this flux line as the microwave control line for C3 and C4 through the weak capacitive coupling. Furthermore, a dc flux penetrating through the superconducting loop is generated with a superconducting coil mounted outside the sample package (see Appendix~\ref{Measurement Setup} for details). This coil is primarily used to maintain the flux bias at the idle bias point, where the smallest \textit{ZZ} interaction between Q1 and Q2 is achieved.

Ignoring the contribution of the resonators and control lines, the system Hamiltonian can be approximated as~\cite{PhysRevApplied.18.034038}
\begin{equation}\label{basic Hamiltonian}
\begin{aligned}
\hat{H}= &~ \frac{4e^2}{2} \hat{\mathbf{n}}^\mathrm{T}  \mathsf{C}^{-1} \hat{\mathbf{n}} -\sum_{i=1}^4  E_{\mathrm{J}i} \cos \hat{\varphi}_i \\
&- E_\mathrm{J5} \cos \left(\hat{\varphi}_4-\hat{\varphi}_3-\varphi_{\mathrm{ex}}\right),
\end{aligned}
\end{equation}
where $e$ represents the elementary charge, $\hat{\mathbf{n}}$ is a vector $\left\{ \hat{n}_i \right\}$, and $\hat{n}_i$ and $\hat{\varphi}_i$ represent the Cooper-pair number operator and phase-difference operator, respectively, with $i \in \{1, 2, 3, 4\}$ corresponding to the nodes of \{Q1, Q2, C3, C4\}. The capacitance matrix $\mathsf{C}$ is defined as $\mathsf{C}_{ii} = \sum_{j=1}^4 C_{ij}$ and $\mathsf{C}_{ij} = -C_{ij}$~($i\neq j$). The reduced flux $\varphi_{\mathrm{ex}} = 2\pi\Phi_\text{ex}/\Phi_0$ denotes the extra phase introduced by the external flux $\Phi_\text{ex}$ divided by the flux quantum $\Phi_0 \equiv h/(2e)$. $E_{\mathrm{J}i} = \Phi_0 I_{\mathrm{c}i}/(2\pi)$ represents the Josephson energy of Josephson junction $i$~(JJ$i$) with the critical current $I_{\mathrm{c}i}$, where $i\in \{1, 2, 3, 4, 5\}$ corresponds to the five junctions shown in the circuit diagram. 


To facilitate an intuitive understanding of the coupling mediated by the DTC, we employ a toy model in which two qubits are coupled via a fixed-frequency transmon (P) and a capacitively shunted flux qubit (CSFQ) (M) [Fig.~\ref{Fig1:wide}(c); see Appendix~\ref{Toy model of the DTC} for the derivation]. The Hamiltonian can be quantized as
\begin{equation}\label{quantized H}
\begin{aligned}
\hat{H} =&~~\omega_1 \hat{a}_1^\dag \hat{a}_1 + \eta_1 \hat{a}_1^\dag \hat{a}_1^\dag \hat{a}_1 \hat{a}_1\\
&+\omega_2 \hat{a}_2^\dag \hat{a}_2 + \eta_2 \hat{a}_2^\dag \hat{a}_2^\dag \hat{a}_2 \hat{a}_2\\
&+\hat{H}_\mathrm{p} + \hat{H}_\mathrm{m}(\varphi_{\mathrm{ex}})\\
&+g_{1\mathrm{p}}\left(\hat{a}_1^\dag \hat{a}_\mathrm{p} + \hat{a}_1 \hat{a}_\mathrm{p}^\dag\right) + g_{2p}\left(\hat{a}_2^\dag \hat{a}_\mathrm{p} + \hat{a}_2 \hat{a}_\mathrm{p}^\dag\right)\\
&+g_{1\mathrm{m}}\left(\hat{a}_1^\dag \hat{a}_\mathrm{m} + \hat{a}_1 \hat{a}_\mathrm{m}^\dag\right) - g_{2m}\left(\hat{a}_2^\dag \hat{a}_\mathrm{m} + \hat{a}_2 \hat{a}_\mathrm{m}^\dag\right), \\
\end{aligned}
\end{equation}
where $\omega_{1(2)}$ and $\eta_{1(2)}$ denote the frequency and anharmonicity of Q1 (Q2), respectively. $\hat{H}_\mathrm{p}$ represents the Hamiltonian of the fixed-frequency transmon~P, while $\hat{H}_\mathrm{m}$ represents the Hamiltonian of the flux-tunable CSFQ~M. The coefficient $g_{i\mathrm{p}(i\mathrm{m})}$ is the capacitive coupling between the qubit Q$i$ and the $\mathrm{p}(\mathrm{m})$-mode, while $\hat{a}$ and $\hat{a}^\dag$ are the lowering and raising operators. The opposite sign of the coupling between the two qubits and the m-mode, suggesting an effective negative capacitance between Q2 and the m-mode.

The energy spectra of the qubits and the DTC are illustrated in Fig.~\ref{Fig1:wide}(d). We denote the diabatic states in the form of $|\mathrm{Q1,}\mathrm{Q2,}\mathrm{P,}\mathrm{M}\rangle$ for the two qubits and the p- and m-modes, which are obtained by dropping all the coupling terms between the four modes in the toy-model Hamiltonian~[Eq.~(\ref{approximation H})]. The adiabatic states with a tilde, $|\reallywidetilde{\mathrm{Q1,}\mathrm{Q2,}\mathrm{P,}\mathrm{M}}\rangle$, are used to indicate hybridized states considering all the couplings and are the eigenstates of the Hamiltonian [Eq.~(\ref{basic Hamiltonian})]. While the adiabatic states provide a precise description of the DTC scheme, the diabatic states provide an intuitive understanding. The simulation results based on the diagonalization of the Hamiltonian [Eq.~(\ref{basic Hamiltonian})] are generated by fine-tuning the designed device parameters to match the experimental data~(see Appendix~\ref{Device parameters} for details). The good agreement between the theoretical model of the DTC and the experimental results confirms its validity. A marginal deviation in the simulation may be attributed to the unconsidered coupling between the transmons and their readout resonators as well as the  control lines.

\section{\textit{ZZ} interaction}\label{ZZ interaction}

The DTC facilitates a high on--off ratio of the longitudinal \textit{ZZ} interaction between the two qubits. By denoting the state energy as $E$, the \textit{ZZ} interaction is formally defined as
\begin{equation}\label{ZZ_interaction}
    \zeta = E_{|\widetilde{1100}\rangle} - E_{|\widetilde{1000}\rangle}  - E_{|\widetilde{0100}\rangle} + E_{|\widetilde{0000}\rangle}.
\end{equation}
The origin of the \textit{ZZ} interaction can be understood through the toy model by accessing the effective coupling between the qubits mediated by the p- and m-modes. Within the dispersive regime of the system, where $|\Delta_{i\mathrm{p}}| = |\omega_{i} - \omega_\mathrm{p}| \gg g_{i\mathrm{p}}$ and $|\Delta_{i\mathrm{m}}| = |\omega_{i} - \omega_\mathrm{m}| \gg g_{i\mathrm{m}}$, the effective coupling between two qubits is given by
\begin{equation}\label{effctive g}
\begin{aligned}
g_{\mathrm{eff}}  = &~\frac{g_{1\mathrm{p}}g_{2\mathrm{p}}}{2}\left(\frac{1}{\Delta_{1\mathrm{p}}} + \frac{1}{\Delta_{2\mathrm{p}}}\right)-\frac{g_{1\mathrm{m}}g_{2\mathrm{m}}}{2}\left(\frac{1}{\Delta_{1\mathrm{m}}} + \frac{1}{\Delta_{2\mathrm{m}}}\right). \\
\end{aligned}
\end{equation}
The contribution of the p-mode can be viewed as a fixed coupling, whereas the tunability of the m-mode frequency results in a variable one. 

The coupling strengths, $g_{i\mathrm{p}}$ and $g_{i\mathrm{m}}$, are approximately equal to each other as the p and m-modes share the same coupling capacitances to the qubits. The effective coupling $g_{\mathrm{eff}}$ can be tuned from negative to positive by adjusting the m-mode frequency across the p-mode frequency with the external flux $\varphi_\mathrm{ex}$. Therefore, $g_{\mathrm{eff}}$ can be tuned to either zero or a specific value, facilitating both negligible \textit{ZZ} interaction for the decoupling and a significantly large one for implementing a CZ gate. While the dispersive approximation works well for the p-mode, the strong coupling between the qubits and m-mode at certain bias flux requires the analysis of interaction between three multi-level modes \cite{PhysRevX.11.021058, chu2021coupler}.

\subsection{Minimum \textit{ZZ} interaction}\label{minimum ZZ interaction} 

\begin{figure}
\includegraphics[scale=1.0]{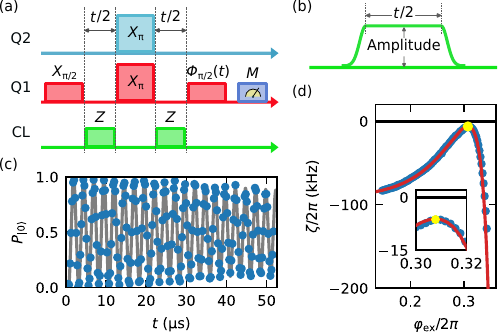}
\caption{\label{Fig2:wide} (a) JAZZ pulse sequence employed for measuring the \textit{ZZ} interaction of the two qubits (Q1 and Q2) with applying the \textit{Z}-pulse to the coupler loop~(CL) of the DTC. The final $\pi/2$ pulse, $\mathit{\Phi}_{\pi/2}(t)$, of Q1 intentionally acquires a phase $\phi$ that is linearly dependent on the interval $t$, providing a baseline frequency to improve the fitting. The sequence considered the margins between the pulses to avoid overlaps, which are not displayed for simplicity. (b) \textit{Z}-pulse shape comprising a flat-top pulse with a duration of $t/2$ as well as smooth Gaussian rise and fall edges with fixed duration. The flux introduced into the coupler loop of the DTC is proportional to the amplitude of the \textit{Z}-pulse. (c) Ground-state population of Q1, $P_{|0\rangle}$, measured at the idle bias point by the JAZZ sequence. The blue experimental data points are fitted with the gray solid line, and its \textit{ZZ} interaction~[$2\pi$$\times$$(-6.3)$~KHz] is extracted and illustrated in (d) with a yellow circle. (d) \textit{ZZ} interaction $\zeta$ as a function of the reduced flux $\varphi_\mathrm{ex}$. The blue data points are experimentally obtained. Simulation results~(red line) are obtained by calculating the energy-level difference using original Hamiltonian~[Eq.~(\ref{basic Hamiltonian})]. The inset provides a zoomed-in view of the \textit{ZZ} interaction near the idle bias point.
}
\end{figure}

We designate the flux bias point that results in the minimum \textit{ZZ} interaction as the idle bias point of the system. This idle bias point serves as the working point for single-qubit gate calibration and the start point for CZ-gate implementation. The flux bias corresponding to the idle bias point is maintained by the current supplied to the superconducting coil.

The \textit{ZZ} interaction is measured using the pulse sequence of the Joint Amplification of \textit{ZZ} interaction (JAZZ) method~\cite{PhysRevLett.125.200504, PhysRevLett.130.260601}. In the JAZZ sequence~[Fig.~\ref{Fig2:wide}(a)], the \textit{ZZ} interaction introduces a relative phase $\phi$ accumulated on the superposition state of Q1, $|0\rangle + e^{i\phi}|1\rangle$, which can be readout from its population measurement, $P_{|0\rangle} = \left(1 - \cos\phi\right)/2$. Therefore, the population is oscillating with the duration $t$, where $t/2$ is defined as the pulse duration of the applied \textit{Z}-pulse. Note that we only sweep the duration of the flat-top of the \textit{Z}-pulse while keeping the duration of the Gaussian rise and fall edges fixed~[Fig.~\ref{Fig2:wide}(b)]. Therefore, the accumulated phase can be simply expressed as $\phi = \zeta t/2 + \phi_0$, where the first term arises from the \textit{ZZ} interaction during the flat-top pulse, and the second term as a fixed phase resulting from the rise and fall edges. Due to the small magnitude of the \textit{ZZ} interaction near the idle bias point, the oscillation of Q1 population measured with the JAZZ sequence becomes relatively slow. To precisely measure the oscillation frequency within a time limited by the qubit coherence time, we intentionally vary the phase of the final $\pi/2$ gate, $\mathit{\Phi}_{\pi/2}(t)$, applied to Q1 linearly with the duration $t$. This introduces an additional relative phase $\omega_\mathrm{b} t$ to the Q1's superposition state, thereby increasing the oscillation frequency with the baseline frequency $\omega_\mathrm{b}$ and leading to improved fitting results. With the frequency $\omega_\mathrm{m}$ extracted from the oscillating signal~[Fig.~\ref{Fig2:wide}(c)], the \textit{ZZ} interaction is calculated as $\zeta = 2(\omega_\mathrm{m} - \omega_\mathrm{b})$. The condition $\omega_\mathrm{b}>|\zeta/2|$ has been chosen to ensure a positive $\omega_\mathrm{m}$, thereby avoiding confusion in the sign of $\omega_\mathrm{m}$ extracted from the oscillating signal with a cosine function.

By varying the amplitude of the \textit{Z}-pulse and measuring its induced oscillation frequency, a flux-dependent \textit{ZZ} interaction can be resolved [Fig.~\ref{Fig2:wide}(d)]. The simulated \textit{ZZ} interaction calculated from the simulated energy levels with~Eq.~(\ref{ZZ_interaction}) is highly consistent with our measurement results. This consistency further confirms the validity of our methods. The minimum \textit{ZZ} interaction is $2\pi \times$$(-6.3)$ kHz, achieved with the external flux bias at $\varphi_\mathrm{ex}/2\pi = 0.309$. With such a small residual \textit{ZZ} interaction at the idle bias point, we have confirmed that it has a negligible influence on the single-qubit gates. Through individual~(simultaneous) randomized benchmarking, we achieved single-qubit gate fidelities of $99.985 \%$~($99.985 \%$) for Q1 and  $99.981 \%$~($99.983 \%$) for Q2, with all fidelity uncertainties below $0.001 \%$ calculated from the fitting error (see Appendix~\ref{Single-qubit gate fidelities} for details).

\subsection{Maximum \textit{ZZ} interaction}\label{maximum ZZ interaction} 

\begin{figure}
\includegraphics[scale=1.0]{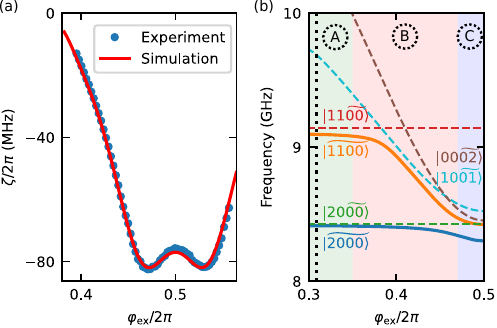}
\caption{\label{Fig3:wide} (a) \textit{ZZ} interaction $\zeta$ measured by the JAZZ pulse sequence around the maximum \textit{ZZ} interaction regime. The blue data points are extracted from the measurements, and the red line is calculated from the simulated energy levels. (b) Simulated energy levels of the higher excited states calculated from the original Hamiltonian~[Eq.~(\ref{basic Hamiltonian})]. The thick colored solid lines show the energy levels when the qubits and coupler transmons are hybridized, while the thin dash lines show the energy levels assuming zero coupling between the qubits and coupler transmons. The vertical dotted line indicates the idle bias point, while the colored regimes labeled with \{A, B, C\} roughly indicate different contribution stages to the \textit{ZZ} interaction.
}
\end{figure}

In contrast to the minimum \textit{ZZ} interaction achieved at the idle bias point, the DTC allows for a significantly larger \textit{ZZ} interaction when $\varphi_\mathrm{ex}/2\pi \sim 0.5$. We applied a similar approach for measuring the larger \textit{ZZ} interaction using the JAZZ protocol. Due to the fast oscillation resulting from the large \textit{ZZ} interaction, the extra oscillation introduced by varying the phase of the final $\mathit{\Phi}_{\pi/2}(t)$ gate is no longer necessary. Instead, we fixed its phase to $90^\circ$ as a simple $Y_{\pi/2}$ gate and extracted \textit{ZZ} interaction from the observed oscillation frequency given as $\zeta = 2 \omega_\mathrm{m}$. The simulation results remain consistent with the measurement results [Fig.~\ref{Fig3:wide}(a)]. A maximum \textit{ZZ} interaction of $2\pi$$\times$$(-82.5)$ MHz is obtained at the external flux $\varphi_\mathrm{ex}/2\pi = 0.47$. By comparing this with the minimum \textit{ZZ} interaction at the idle bias point, we obtain an on--off ratio of $1.3 \times 10^4$.

When considering the subspace comprised solely of qubits, given the toy model in the dispersive regime, the energy-level repulsion between $|02\rangle$ and $|11\rangle$, akin to the single-transmon-coupler case, contributes to the \textit{ZZ} interaction. However, it is not applicable in the strong coupling regime, where larger \textit{ZZ} interactions arises and a coupler-state-mediated process should be considered \cite{chu2021coupler}. We denote the states $|\mathrm{Q1,}\mathrm{Q2,}\widetilde{\mathrm{P,}\mathrm{M}}\rangle$ as eigenstates numerically calculated from the Hamiltonian~[Eq.~(\ref{basic Hamiltonian})] by suppressing the qubit--coupler couplings, for the purpose of the following explanation. Starting from the idle flux point, the energy $E_{|\widetilde{1100}\rangle}$ decreases with increasing the flux bias because of the interaction between the diabatic states, $|11\widetilde{00}\rangle$ and $|10\widetilde{01}\rangle$~[Fig.~\ref{Fig3:wide}(b)]. However, this does not lead to a large \textit{ZZ} interaction because the energy $E_{|\widetilde{0100}\rangle}$ also similarly decreases given the interaction between $|01\widetilde{00}\rangle$ and $|00\widetilde{01}\rangle$~[region A in Fig.~\ref{Fig3:wide}(b)]. With further increasing the flux bias, the contribution of the large \textit{ZZ} interaction arises due to the large repulsion between $|00\widetilde{02}\rangle$ and $|10\widetilde{01}\rangle$, which induces extra decrease of $E_{|\widetilde{1100}\rangle}$ mediated by $|10\widetilde{01}\rangle$~[region B in Fig.~\ref{Fig3:wide}(b)]. A slight decrease of the \textit{ZZ} interaction magnitude in the range of $0.47 < \varphi_\mathrm{ex}/2\pi < 0.5$ can be attributed to the repulsion from $|20\widetilde{00}\rangle$ and $|00\widetilde{02}\rangle$ to $|10\widetilde{01}\rangle$, consequently increasing $E_{|\widetilde{1100}\rangle}$~[region~C in Fig.~\ref{Fig3:wide}(b)]. 


Such a large \textit{ZZ} interaction is crucial for the implementation of a fast CZ gate for superconducting qubits. However, taking into account the highly complicated energy levels with multiple anti-crossings, a finely tuned \textit{Z}-pulse should be provided for an adiabatic process to suppress the leakage error.

\section{CZ-gate implementation}\label{CZ gate implementation}

\subsection{Preliminaries for CZ gate}

Firstly, let us emphasize the conditions of the idle bias point that are prepared before implementing the CZ gate. A persistent flux is introduced into the superconducting loop of the DTC, generated by the superconducting coil. The idle bias point is confirmed through the characterization of the JAZZ sequence, which gives the near-zero minimum \textit{ZZ} interaction magnitude. Then, the single-qubit gates for the two qubits are precisely calibrated with the DRAG method~\cite{PhysRevLett.103.110501, PhysRevA.82.042339}. We note that a CZ gate with high fidelity is realized through precise calibration of the CPHASE gate exploiting the \textit{ZZ} interaction. Therefore, single-qubit gates with high fidelities are essential, as they otherwise would introduce extra phases that could miscalibrate the CZ-gate phase during optimization. 

Before the CZ-gate implementation, it is necessary to carefully calibrate the distortion of a \textit{Z}-pulse, which arises during the transmission along the fast flux line. However, the distortion characterization becomes tricky if there is no readout resonator for coupler's phase tomography. The DTC scheme shows its advantages in addressing this issue. Firstly, the single tunable element among the four transmons simplifies the characterization process compared to the STC scheme, which usually consists of a tunable coupler and at least another tunable qubit~\cite{PhysRevX.11.021058}. Secondly, due to the large coupling and level repulsions between the qubits and coupler transmons, we can use the qubits to characterize the distorted \textit{Z}-pulse that is introduced into the DTC loop. A Ramsey-type experiment along with phase tomography~\cite{PhysRevX.11.021058}, typically employed for pulse-distortion characterization, is detailed in Appendix~\ref{$Z$-pulse distortion calibration}. Subsequently, a predistortion of the \textit{Z}-pulse is applied during the optimization and benchmarking of the CZ gate.

\subsection{CZ-gate benchmarking}\label{CZ gate benchmarking} 

\begin{figure*}
\includegraphics[scale=1.0]{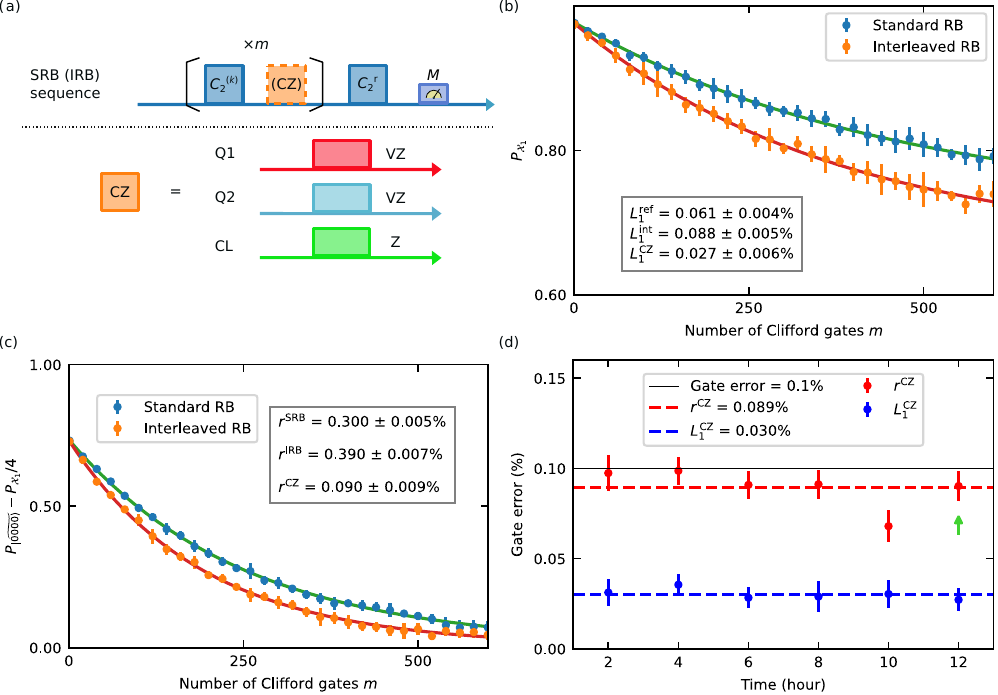}
\caption{\label{Fig4:wide} \Add{(a)~Pulse sequence for the randomized benchmarking~(RB). It encompasses $m$ randomly selected two-qubit Clifford gates~($C_2^{(k)}$; $k=1,\ldots, m$), with and without the CZ gates interleaved, concluding with a recovery Clifford gate~($C_2^\mathrm{r}$), denoted as standard RB~(SRB) and interleaved RB~(IRB), respectively. The entire sequence collectively functions as an identity operation applied to the two qubits. The population of $|\widetilde{0000}\rangle$, $P_{|\widetilde{0000}\rangle}$, is finally measured ($M$) as the sequence fidelity. The population $P_{\mathcal{X}_1}$ in the computational subspace, $\mathcal{X}_1 = \{|\widetilde{0000}\rangle, |\widetilde{0100}\rangle, |\widetilde{1000}\rangle, |\widetilde{1100}\rangle\}$, is also measured for leakage error benchmarking. The optimized CZ gate comprises a \textit{Z}-pulse with 48-ns duration applied to the coupler loop of the DTC and two virtual Z (VZ) gates applied to the two qubits~(see Appendix~\ref{CZ gate optimization by RL algorithm} for details). (b)~Leakage RB~(LRB). Each data point and its error bar are the average and standard deviation of measured results on 10 randomly selected RB sequences illustrated in~(a), respectively. Both the reference and interleaved results are fitted with exponential curves. (c)~Standard and interleaved RB results characterizing the CZ-gate error. The error rates, $r^\mathrm{SRB}$ and $r^\mathrm{IRB}$, and the CZ-gate error $r^\mathrm{CZ}$ are calculated from the exponential fitting based on Eq.~(7). (d)~12-hour timeframe measurement of the CZ-gate error $r^\mathrm{CZ}$ and corresponding leakage error $L_1^\mathrm{CZ}$, where the red and blue dashed lines depict the respective average errors. The last data point (pointed by a green arrow) of the CZ-gate error is calculated from the results in~(c).}
}
\end{figure*}

The \textit{ZZ} interaction can be efficiently activated with an applied \textit{Z}-pulse applied through the DTC loop, resulting in a  CPHASE gate defined as $U_{\mathrm{CPHASE}}=\operatorname{diag}\left(e^{\mathrm{i} \theta_0}, e^{\mathrm{i} \theta_1}, e^{\mathrm{i} \theta_2}, e^{\mathrm{i} \theta_3}\right)$. A CZ gate is achieved with a $\pi$ phase accumulated, i.e., $\theta_{\mathrm{CZ}}=\theta_3-\theta_1-\theta_2+\theta_0 = \pi$. Though each of the two qubits contains only a single Josephson junction, usually considered as a fixed-frequency qubit, the substantial repulsion of their energy levels from the coupler transmons leads to a dynamical change of their frequencies during the CZ gate. Therefore, the inclusion of a commonly used error-free virtual Z (VZ) gate~\cite{PhysRevA.96.022330} remains essential in the DTC scheme to compensate for the additional $Z$-rotation~($\theta_1$ or $\theta_2$) of each qubit [Fig.~\ref{Fig4:wide}(a)]. 

To optimize the CZ gate, we started with a Slepian pulse shape~\cite{press2007numerical} as an initial guess and employed an optimization process driven by the RL algorithm, following the guidelines outlined in Ref.~\citenum{PhysRevX.12.011059}. While the \textit{Z}-pulse shape and VZ gates can be jointly optimized together~\cite{PhysRevX.13.031035}, we chose to decouple them into two parts to enhance the efficiency of the optimization process (see Appendix~\ref{CZ gate optimization by RL algorithm} for details). The decoupled optimization process is pivotal for the DTC scheme, considering the high sensitivity of the VZ-gate phase to the shape of the \textit{Z}-pulse. After those meticulous optimizations, we confirmed the implemented gate as a CZ gate through quantum process tomography (see Appendix~\ref{Quantum process tomography} for details). The achieved fidelity was $\sim$95.9\%, primarily affected by the state-preparation-and-measurement (SPAM) error~\cite{PhysRevLett.119.180501, PhysRevLett.109.080505}.

A pure CZ-gate fidelity can be accessed through two-qubit standard and interleaved randomized benchmarking (SRB and IRB)~\cite{PhysRevLett.109.080505, PhysRevA.87.030301,  barends2014superconducting}. \Add{A schematic of the SRB~(IRB) sequence is illustrated in Fig.~\ref{Fig4:wide}(a), where the survival probabilities, $P_{|\widetilde{0000}\rangle}$ and $P_{\mathcal{X}_1}$, are measured against the number $m$ of the randomly selected Clifford gates with~(without) CZ gates interleaved in the sequence. The total population in the computational subspace, $P_{\mathcal{X}_1}$, is defined as
\begin{equation}
P_{\mathcal{X}_1}=P_{|\widetilde{0000}\rangle}+P_{|\widetilde{0100}\rangle}+P_{|\widetilde{1000}\rangle}+P_{|\widetilde{1100}\rangle},
\end{equation}
which is used for leakage randomized benchmarking~(LRB)~\cite{PhysRevLett.123.120502} and is fitted by
\begin{equation}
P_{\mathcal{X}_1}(m)=A + B \lambda_L^{m},
\end{equation}
as in Fig.~4(b).
In the presence of leakage, the survival probability $P_{|\widetilde{0000}\rangle}$ estimated from the RB should be modeled using double exponential decays~\cite{PhysRevA.97.032306}, which usually makes the fitting unreliable. To address this, we derived a single exponential decay model, fitting the results with
\begin{equation}\label{P00-fit}
    P_{|\widetilde{0000}\rangle}(m) - P_{\mathcal{X}_1}(m)/d = C\lambda_r^{m} +D,
\end{equation}
where $d = 4$ is the dimension of the computational subspace~[Fig.~\ref{Fig4:wide}(c)]~(See Appendix~\ref{CZ gate evaluation} for details).
We note that the CZ-gate infidelity comprises two contributions, the leakage error $L_1^\mathrm{CZ}$ estimated from $\lambda_L$ and the gate error $r^\mathrm{CZ}$ estimated from $\lambda_r$.}

\Add{
Firstly, the leakage error $L_1^\mathrm{SRB(IRB)}$ is estimated by
\begin{equation}
L_1^\mathrm{SRB(IRB)}= (1 - A)(1 - \lambda_L^\mathrm{SRB(IRB)}).
\end{equation}
With the estimated reference leakage error $L_1^\mathrm{SRB}$ and the interleaved one $L_1^\mathrm{IRB}$, the leakage error of the CZ gate can be calculated as
\begin{equation}
    L_1^{\mathrm{CZ}}=1-\frac{1-L_1^\mathrm{IRB}}{1-L_1^\mathrm{SRB}}.
\end{equation}
}

\Add{
On the other hand, the error rate $r^\mathrm{SRB(IRB)}$ in standard~(interleaved) RB is defined (in a similar manner to the widely used methods~\cite{PhysRevX.11.021058, PhysRevX.13.031035, PRXQuantum.4.010314}) as
\begin{equation}
    r^\mathrm{SRB(IRB)} = (1-\lambda_r^\mathrm{SRB(IRB)})(1-1/d).
\end{equation}
The CZ-gate error can be obtained as
\begin{equation}
    r^\mathrm{CZ} =1-\frac{1-r^\mathrm{IRB}}{1-r^\mathrm{SRB}}\approx \frac{(d-1)}{d}\left(1-\frac{\lambda_r^\mathrm{IRB}}{ \lambda_r^\mathrm{SRB}}\right) .
\end{equation}
A typical result of LRB with the estimated leakage error $L_1^\mathrm{CZ}$ of $0.027 \pm 0.006\%$ is depicted in Fig.~\ref{Fig4:wide}(b), along with an evaluation of the CZ-gate error of   $r^\mathrm{CZ} = 0.090 \pm 0.009\%$ in Fig.~\ref{Fig4:wide}(c). 
}

To validate the stability of our results, we conducted the measurements of the CZ-gate error within a 12-hour timeframe [Fig.~\ref{Fig4:wide}(d)]. Throughout the measurement, the CZ-gate parameters are only optimized once at the beginning, while the single-qubit gates are repeatedly calibrated after each benchmarking sequence measurement. The six measured CZ-gate errors \Add{$r^\mathrm{CZ}$}, all below 0.1\%, underscore the reliability and stability of our achievement. The average result gives the \Add{$r^\mathrm{CZ}$ of $0.09 \pm 0.01\%$, while the leakage error $L_1^\mathrm{CZ}$~\cite{PhysRevA.97.032306, PhysRevLett.123.120502} is also evaluated each time with an average of $0.030 \pm 0.003 \%$}.

\Add{Here, we note that the CZ-gate error $r^\mathrm{CZ}$ introduced above is not equivalent to the gate infidelity in the presence of leakage error. The total CZ-gate infidelity ($1-\bar{F}$) encompasses both $r^\mathrm{CZ}$ and the leakage error $L_1^\mathrm{CZ}$ measured by LRB. Therefore, the total gate fidelity $\bar{F}$ is given as~\cite{PhysRevA.97.032306}
\begin{equation}
    \bar{F}=1 - \frac{L_1^\mathrm{CZ}}{d} - r^\mathrm{CZ},
\end{equation}
which is evaluated to be $99.90\pm 0.01\%$ in our experiment. 
}

\Add{
Note that $r^\mathrm{CZ}$ also incorporates contributions from leakage error. This inclusion is based on the assumption of a depolarizing leakage model for the qubit system, where $\lambda_r = (1-L_1)(1-p_D)$ for Eq.~(\ref{P00-fit}), with $p_D$ representing the depolarizing rate in the computational subspace~\cite{PhysRevA.97.032306}. Under this model, the total CZ-gate fidelity can also be expressed as
\begin{equation}
    \bar{F}=1 - L_1^\mathrm{CZ} -  \frac{d - 1}{d}p_D^\mathrm{CZ},
\end{equation}
with the assumption of $L_1^\mathrm{CZ}, p_D^\mathrm{CZ} \ll 1$~(See Appendix~\ref{CZ gate evaluation} for details).
This formulation provides a more explicit representation of the gate infidelity, delineating contributions from the leakage and depolarizing effects. Besides, it also provides the depolarizing-induced gate error $r_D^\mathrm{CZ}$:
\begin{equation}\label{depolarizing error}
    r_D^\mathrm{CZ} \equiv \frac{d - 1}{d}p_D^\mathrm{CZ} = r^\mathrm{CZ} - \frac{d - 1}{d}L_1^\mathrm{CZ}.
\end{equation}
}

\subsection{CZ-gate error analysis}\label{CZ-gate error analysis}

\begin{figure}
\includegraphics[scale=1.0]{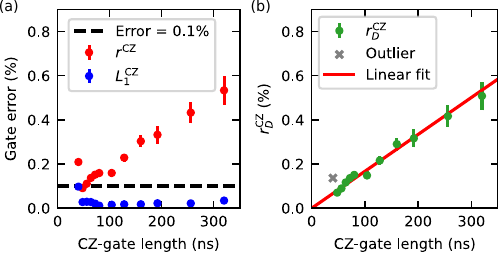}
\caption{\label{Fig5:wide} \Add{(a) CZ-gate error $r^\mathrm{CZ}$ and the corresponding leakage error $L_1^\mathrm{CZ}$ measured and calculated from the RB results for various CZ-gate lengths. The error bars are calculated from the fitting error of the RB results. The error bars of the leakage error data are smaller than the symbol size. (b) Depolarizing-induced gate error $r_D^\mathrm{CZ}$ calculated with Eq.~(\ref{depolarizing error}). The data points and their standard deviations are calculated from the data in (a). The red line shows a linear fit. The cross marker is an obvious outlier, corresponding to a gate length of 40 ns, and is therefore excluded from the linear fitting.}
}
\end{figure}

The CZ-gate errors are typically categorized into two parts: coherent and incoherent errors. Coherent errors primarily encompass non-adiabatic process-induced leakage error and imperfect CZ- and VZ-phase errors. Though difficult, these errors can be minimized through precise tuning of the control pulse. On the other hand, incoherent errors, arising from limited qubit coherence due to various noise sources, present a more challenging mitigation task. The leakage error can be directly measured simultaneously with the LRB experiment~\cite{PhysRevX.11.021058}, and a coherent error rate $<$0.01\% due to the finely calibrated phase error $<$1.5\textdegree~can be considered negligible at current stage~\cite{PhysRevX.13.031035}. Even with considerable efforts, however, estimating incoherent errors for high-fidelity two-qubit gates remains challenging~\cite{PhysRevLett.125.240503, PhysRevLett.127.080505, PhysRevX.11.021058, PhysRevX.13.031035}. Single-qubit coherence times are typically characterized to elucidate the incoherent error of a two-qubit gate. However, compared to the randomized benchmarking measurement, the separate decoherence measurements often underestimate the decoherence experienced by the qubits when the two-qubit gate is applied. Despite those imperfect error-budget estimations, they have provided guidance for enhancing gate fidelity to surpass the threshold~\cite{fowler2012surface, arute2019quantum} required for implementing the surface code~\cite{bravyi1998quantum, dennis2002topological} over the past decades. Therefore, we present here only our rough estimation of the contributors to the CZ-gate error and place more emphasis on the insights gained from the results.

The CZ-gate errors, with varying gate length, were measured using the CZ gate optimized with the same methods described above, simultaneously with the assessment of leakage error~[Fig.~\ref{Fig5:wide}(a)]. The leakage error are relatively small~(0.010--0.035 \%), except for the one about $0.1\%$ with the CZ-gate length of 40 ns, as illustrated by the leftmost point in Fig.~\ref{Fig5:wide}(a). We attribute the sudden increase of the leakage error to the failure of optimizing the \textit{Z}-pulse shape in experiment for such a short gate length. This is supported by achieving a CZ-gate fidelity higher than 99.99\% in numerical simulation by using a theoretically optimized pulse shape with gate length of 40 ns. The discrepancy could be attributed to the bandwidth limitation of the electric devices and circuits, and/or the failure of the RL-based optimization. We also observed a slight increase in the leakage error with gate length exceeding 100 ns. This is attributed to the fact that the contribution of the leakage error becomes much smaller compared to other errors, making it challenging to optimize precisely during the RL-based optimization process.


\Add{We model the depolarizing-induced error $r_D^\mathrm{CZ}$ as the sum of incoherent errors (see Appendix~\ref{incoherent error}) and a gate-length-independent error $r_0$:
\begin{equation}
    r_D^\mathrm{CZ} =  r_\text{incoherent} + r_0 = \frac{2}{5} \frac{t_{\text{gate}}}{T_{\text{eff}}} + r_0,
\end{equation}
where $t_{\text{gate}}$ is the CZ-gate length and $T_\mathrm{eff}$ represents the effective coherence time experienced by the entire system during the IRB experiment. We fit the gate-length dependence of the depolarizing-induced error with the linear model~[Fig.~\ref{Fig5:wide}(b)], except for the data marked as an obvious outlier. We attribute this outlier to a sudden drop in qubit coherence, which could be caused by the significantly stronger hybridization between the qubit and coupler with the requirement of such a short gate length. The high Pearson correlation coefficient ($\approx 0.995$) for the linear model, with a small $p$-value ($p$~$<$~$10^{-9}$), indicates that the correlation is statistically significant. The negligibly small offset we obtained, $r_0 = -0.0007 \pm 0.008\%$, validates our assumption of the depolarizing error model.}

In principle, the effective coherence time $T_\mathrm{eff}$ mainly encompasses both $T_1$ and $T_\phi$ of the two qubits as well as the influence of the decoherence in the two coupler transmons, given they are highly coupled during the CZ gate. The linear relation is modeled with neglecting the 1/\textit{f} flux noise as it only plays a marginal role~\cite{PhysRevX.11.021058}~(see Appendix~\ref{incoherent error} for details). Note that the effective coherence time, \Add{$T_{\text{eff}}^{\mathrm{exp}} = 23.9\pm 1.5$~\textmu s}, cannot be entirely predicted from the coherence time of the two individual qubits measured at the idle bias point, $T_{\text{eff}}^{\mathrm{est}} = 67.6\pm 11.4$~\textmu s, similarly to other works~\cite{PhysRevLett.127.080505, PhysRevX.13.031035}. We attribute this to the possibility of additional decoherence introduced when the CZ gate is applied. \Add{We acknowledge that purity benchmarking can be employed to differentiate between coherent and incoherent errors~\cite{PhysRevApplied.6.064007}. However, the requirements for full state tomography, including states in the non-computational subspace, as well as the explicit error budget when leakage error emerge, remain open questions that extend beyond the scope of this work.}

\section{conclusion and discussion}\label{conclusion and discussion}

In this work, our primary demonstration focused on the DTC scheme, showcasing its high-performance experimentally. \Add{We began by presenting an intuitive model that describes the coupling mediated by the DTC. The toy model told us that Josephson junctions JJ3 and JJ4 could potentially be replaced by linear inductors to realize near-zero ZZ interaction at the idle bias point, while the nonlinearity of JJ5 must be taken into account to provide the tunable coupling strength and enable the emergence of large ZZ interactions.} We also achieved simultaneous high fidelity and stability in both single- and two-qubit gates. This achievement was made possible by the finely designed and fabricated transmons with high coherence, as well as a large on--off ratio of the qubit--qubit coupling achieved through carefully selected circuits parameters. With meticulous calibration, we effectively ``turned off'' the coupling by tuning the flux into the DTC loop to the idle bias point. Subsequently, we dynamically activated the coupling with an optimized \textit{Z}-pulse passing through the fast flux line. We opted for a CZ-gate length of 48 ns, striking a balance between leakage error and incoherent error, and achieved a CZ-gate fidelity of \Add{$99.90 \pm 0.01\%$} stably during a 12-hour measurement.

To advance toward even higher gate fidelity, there remain issues that require thorough investigation and complete understanding. Firstly, a shorter gate length for the CZ gate is preferred to mitigate the incoherent error. However, a faster CZ gate requires larger coupling capacitance, which may degrade qubit coherence because of its stronger coupling to the coupler transmons. Besides, another challenge lies in addressing the incapability to optimize such a short-duration pulse effectively for suppressing the leakage error.  Secondly, there exists additional decoherence involved during the benchmark of the CZ gate, which may be due to multi-level mixing of the total system and extra noise channels activated during the \textit{Z}-pulse. 

We note that high-performance gate operation is not the sole metric for evaluating the quality of a qubit coupling scheme. The DTC scheme brings various other advantages, such as less frequency-collision probability due to highly detuned qubits, flexible spatial arrangement of qubits, and simplified control degrees of freedom related to the single tunability element. These attributes make it highly promising as a candidate for implementing NISQ applications and quantum error correction in the near future.

\begin{acknowledgments}
We wish to acknowledge K. Kusuyama and Y. Sakoda for the Ta-film deposition; A. Badrutdinov for the TSV\nobreakdash-fabrication assistance. This research was partly funded by the Ministry of Education, Culture, Sports, Science and Technology~(MEXT) Quantum Leap Flagship Program~(Q-LEAP) (Grant No.~JPMXS0118068682).

\end{acknowledgments}

\appendix

\section{Fabrication}\label{Fabrication}

The fabricated sample chips are based on tantalum-film-based superconducting circuits~\cite{place2021new, wang2022towards}, which has been demonstrated to showcase an extended coherence time. In the fabrication process, a tantalum (Ta) film of around 135\nobreakdash-nm thick is initially sputtered on a pre-cleaned high-resistivity~($>$10~k$\Omega$$\cdot$cm) (100)-oriented silicon wafer at 300$^\circ$C. The growth of $\alpha$-Ta film is simply confirmed by the resistivity of $\sim$15~$\mu \Omega$$\cdot$cm~\cite{javed2010investigation} at ambient condition. Then, the resonators, qubit capacitances, and control lines are patterned through photolithography. After development, the exposed Ta film is subjected to reactive ion etching employing CF$_4$ gas~\cite{wang2022towards}. Following the wafer cleaning procedure involving organic remover, oxygen plasma, and hydrofluoric acid, we create Dolan-bridge-type Josephson junctions through aluminum (Al) deposition and lift-off, employing electron-beam lithography~\cite{dolan1977offset}. Subsequently, through-silicon vias (TSV) are produced using the Bosch process and are further metalized through Al deposition and lift-off, employing an additional photolithography step. Finally, the wafer is diced into 2.5$\times$5-mm$^2$ chips, which are wire-bonded to a home-designed printed circuit board~(PCB).

\section{Measurement setup}\label{Measurement Setup}

\begin{figure}
\includegraphics[scale=1.0]{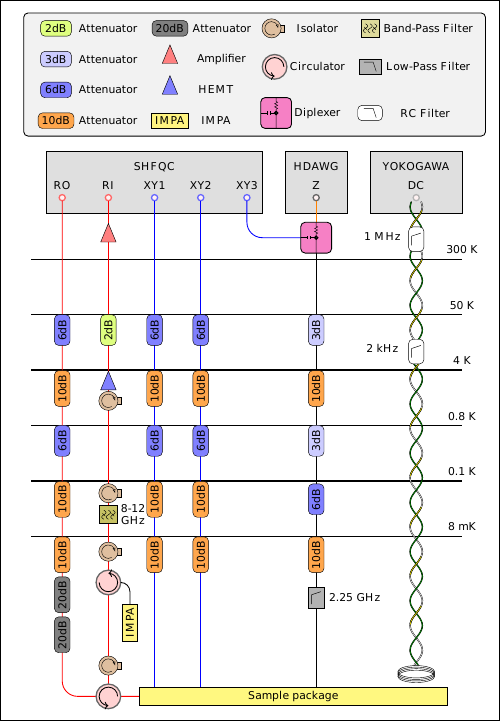}
\caption{\label{FigS1:wide} Schematic diagram of the measurement setup. In the device SHFQC, the RO and RI ports are used to generate and receive the microwave tones for qubit readout, respectively.
}
\end{figure}

Figure~\ref{FigS1:wide} depicts a schematic diagram illustrating the measurement setup. Under ambient conditions, two arbitrary waveform generators~(AWGs) are employed. One AWG~(Zurich Instruments SHFQC) stands out as it has the capability to directly generate signals up to 8.5 GHz without the requirement for additional mixers. This feature is leveraged for producing DRAG waveforms essential for qubit control. Moreover, the device excels in generating and demodulating multiple-frequency signals, facilitating the simultaneous multiplexed readout of the three resonators. Another AWG device~(Zurich Instruments HDAWG8) is employed for generating the \textit{Z}-pulse. It incorporates real-time precompensation to effectively mitigate distortion in the \textit{Z}-pulse waveform. A dc source~(Yokogawa GS200) is employed to supply a constant dc current to the superconducting coil, which is crucial for maintaining a stable and consistent flux bias in the coupler loop. All of these electronic devices are synchronized through a common rubidium frequency standard~(Stanford Research Systems FS725), ensuring precise signal phase alignment.

For qubit readout and control, the waveforms are conveyed through coaxial cables, which are interconnected by a series of attenuators~(XMA) positioned at various temperature stages upon reaching the base temperature of $\sim$8~mK within the dilution refrigerator (BlueFors LD400). The microwave tone for qubit readout is transmitted through a circulator and subsequently reflected by the sample chip. Following this, the signal carrying the qubit information undergoes another transmission through the same circulator. Once it passes through an isolator, the signal enters a home-built impedance-matched Josephson parametric amplifier~(IMPA)~\cite{roy2015broadband} via an additional circulator in the process. The pump and bias setups for the IMPA are omitted in the figure for the sake of simplicity. After being amplified and reflected by the IMPA, the signal is transmitted into the second circulator and redirected toward a sequence of isolators. Amplified at 4 K by a high-electron-mobility-transistor~(HEMT) amplifier~(LNF LNC4\_16C) and at 300 K by a low-noise microwave amplifier, the signal is then received and demodulated for qubit state discrimination. Each of the two qubits has its dedicated \textit{XY} drive lines, while the control signals for \textit{XY} and $Z$ of the coupler transmons converge onto the fast flux line through a diplexer~(QMC 0218LNM) before entering the fridge. To mitigate thermal excitation of the coupler modes while maintaining efficient control of \textit{XY} and $Z$ for coupler transmons, a low-pass filter~(Mini-Circuits VLF-2250+) is employed. All these signals are transmitted through the cables into the package, which serves as the enclosure for holding the chip. Additionally, a dc current is applied to bias the DTC's loop, transmitted through a twisted pair line into a superconducting coil outside the sample package. To mitigate high-frequency noise, a homemade low-pass filter~($<$1.6~kHz) at the 4-K stage and a $\pi$-filter~($<$1~MHz) at room temperature are implemented. To safeguard the qubit from environmental radiation~\cite{gordon2022environmental}, the sample chip and its PCB are initially enclosed within the copper-made sample package and further shielded magnetically with an aluminum can and a two-layer $\mu$-metal shield at the base temperature.

\section{Device parameters}\label{Device parameters}

\begin{table}
\caption{Circuit parameters for the simulation.}
    \centering
    \begin{tabular}{cccccc}
        \hline\hline \text { Capacitance }& & \text { $C_{11}$ } & \text { $C_{22}$ } & \text { $C_{33}$ } & \text { $C_{44}$}  \\
        \hline \text { Value (fF) } & & 91.86 & 91.79 & 110.27 & 106.36\\
        \hline\hline
    \end{tabular}

     \begin{tabular}{ccccccc}
        \hline\hline \text { Critical current }& & \text { $I_{\mathrm{c}1}$ } & \text { $I_{\mathrm{c}2}$ } & \text { $I_{\mathrm{c}3}$ } & \text { $I_{\mathrm{c}4}$} & \text { $I_{\mathrm{c}5}$}  \\
        \hline \text { Value (nA) } & & 26.13 & 31.93 & 47.73 & 47.68&10.32\\
        \hline\hline
    \end{tabular}
    
    \begin{tabular}{cccccccc}
        \hline\hline \text { Capacitance }& & \text { $C_{12}$ } & \text { $C_{13}$ } & \text { $C_{14}$ } & \text { $C_{23}$} & \text { $C_{24}$} & \text { $C_{34}$}  \\
        \hline \text { Value (fF) } & & 0.04 & 5.73 & 0.17 & 0.26 & 5.77 & 1.73 \\
        \hline\hline
    \end{tabular}
    \label{tab:circuit_params}
\end{table}

\begin{table}
\caption{Dressed-state frequencies and anharmonicities of the transmons and readout resonators at the idle bias point. P and M stands for the plus- and minus-sign composite modes of the coupler transmons, behaving like a fixed-frequency transmon and a CSFQ, respectively.}
    \centering
    \begin{tabular}{cccccc}
        \hline\hline \text { Qubit }& & \text { Q1 } & \text { Q2 } & \text { P } & \text { M}  \\
        \hline \text {Readout frequency (GHz) } & & 8.184 & 8.261 & 8.101 & 8.101\\
        \text {Drive frequency (GHz) } & & 4.314 & 4.778 & 5.495 & 5.373 \\
        \text {Anharmonicity (MHz) } & & $-212$ & $-199$ &  & \\
        \hline \hline
    \end{tabular}
    \label{tab:qubit_params}
\end{table}

\begin{figure}
\includegraphics[scale=1.0]{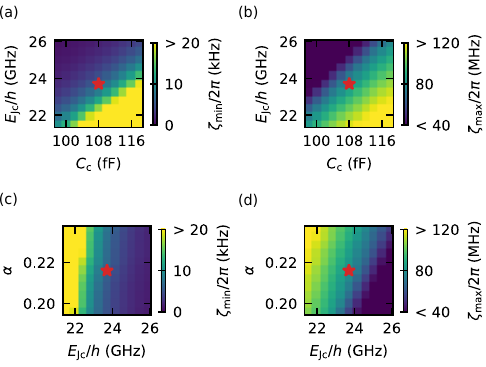}
\caption{\label{FigS12:wide} Numerical simulation of the \textit{ZZ} interaction with fixed parameters of $\alpha = 0.216$ for~(a) and~(b) and $C_\mathrm{c} = 108$ fF for~(c) and~(d). The minimum and maximum \textit{ZZ} interactions, $\zeta_\mathrm{min}$ and $\zeta_\mathrm{max}$, are obtained with different flux biases through the DTC loop for the idle bias point and fast CZ gate, respectively. The red  stars shown in the figures indicate the parameters used for the experiment.
}
\end{figure}

The circuit parameters (Table~\ref{tab:circuit_params}) are used for simulating the energy levels and \textit{ZZ} interaction of the chip used in this work. The frequencies and anharmonicities of the transmon and resonator dressed states~(Table~\ref{tab:qubit_params}) are measured at the idle bias point with~$\varphi_\mathrm{ex}/2\pi =0.309$. We note that the frequency detuning~($\Delta_{21}/2\pi = 464$ MHz) between the two qubits is larger than their anharmonicities, indicating that they are positioned outside of the straddling regime.

Given the complicated energy-level hybridization, the \textit{ZZ} interaction can only be precisely calculated through numerical simulation. Therefore, we manually search the circuit parameters to achieve an acceptable minimum \textit{ZZ} interaction at the idle bias point, as well as a sufficiently large \textit{ZZ} interaction for a fast CZ gate. The following steps describe our procedure, where the presence of readout resonators and control lines are neglected because of their marginal influence on the \textit{ZZ} interaction between qubits.
\begin{enumerate}
    \item The frequencies of the two data qubits are chosen to be within 4--5.5~GHz, with anharmonicities around $-200$~MHz to enable fast single-qubit gates. A detuning of 500~MHz, which is larger than their anharmonicities, is used to demonstrate qubit operation outside the straddling regime.
    \item Stray capacitances, such as $C_{14}$, $C_{23}$, and $C_{12}$, are usually very small ($<$1 fF) and contribute minimally to the \textit{ZZ} interaction. So, they can be neglected. The mutual capacitance $C_{34}$, though small, is comparable to the coupling capacitance and is usually considered as a fixed value during the search for other circuit parameters. This holds true even with minor changes to the structures around the DTC loop, where $C_{34}$ arises. The stray capacitance values can be obtained using finite element analysis with commercial software like Sonnet or COMSOL.
    \item The coupling capacitances $C_g = C_{13} = C_{24}$ are chosen to be 5\% to 10\% of the data qubit shunt capacitance, corresponding to a 100--200-MHz coupling between the data and coupler transmons empirically. A larger coupling capacitance results in a larger \textit{ZZ} interaction, facilitating faster CZ gates. However, it permits only small variations in other parameters to achieve minimal \textit{ZZ} interaction at the idle bias point. For the following simulation, we fixed $C_{13} = C_{24} = 5.7$ fF as an example. It should be noted, however, that this value should be finely tuned for optimal circuit parameters.
    \item The ratio $\alpha = E_{\mathrm{J5}}/E_{\mathrm{J3}} = E_{\mathrm{J5}}/E_{\mathrm{J4}}$ is first fixed to search for the circuit parameters $C_{\mathrm{c}} \equiv C_{33} = C_{44}$ and $E_{\mathrm{Jc}} \equiv E_{\mathrm{J3}} = E_{\mathrm{J4}}$, as demonstrated in~Figs.~\ref{FigS12:wide}(a) and~(b). We use $\alpha = 0.216$, the same as in our device, which should not be too large ($<$0.25, empirically). Intuitively, a larger $C_{\mathrm{c}}$ and smaller $E_{\mathrm{Jc}}$ lead to lower coupler--mode frequencies, resulting in stronger coupling between the data qubits and the DTC. This induces larger \textit{ZZ} interactions for both the minimum and maximum \textit{ZZ} interactions, $\zeta_\mathrm{min}$ and $\zeta_\mathrm{max}$, obtained by varying the flux through the DTC. A parameter set of~\{$C_{\mathrm{c}}$, $E_{\mathrm{Jc}}$\} is chosen to balance $\zeta_\mathrm{min}$ at the idle bias point and $\zeta_\mathrm{max}$ for achieving a fast CZ gate.
    \item We fix the value of $C_{\mathrm{c}}$ obtained in step 4 and search for the parameter set \{$E_{\mathrm{Jc}}, \alpha$\}, as shown in Figs.~\ref{FigS12:wide}(c) and~(d). A larger $\alpha$ induces a stronger coupling between the two coupler transmons, which indirectly increases the coupling between the two data qubits and their \textit{ZZ} interaction. Therefore, these parameters should also be determined to balance the minimum and maximum \textit{ZZ} interactions, similar to step 4.
    
\end{enumerate}

The simple but effective procedure shown above demonstrates our search for the parameter set \{$C_g$, $C_{\mathrm{c}}$, $E_{\mathrm{Jc}}$, $\alpha$\} of the DTC. Steps 3--5 can be repeated several times to obtain an optimal set. A direct parameter search in the four-dimensional space would be possible but time-consuming. Using our procedure, a desired parameter set that is feasible with current fabrication techniques can be easily obtained. We note that these values, shown in Table~\ref{tab:circuit_params}, are slightly different from the theoretically designed ones, which is due to the fabrication error.

\section{Toy model of the DTC}\label{Toy model of the DTC}

\subsection{Toy model without qubits}\label{Approximation model of DTC}

We first derive a toy model of the DTC without considering the data qubits. Assuming the symmetric design of $C_{33} = C_{44} = C_\mathrm{c}$ and $E_{\mathrm{J}3} = E_{\mathrm{J}4} = E_{\mathrm{J}}$ as well as denoting $E_{\mathrm{J}5} = \alpha E_{\mathrm{J}}$~($\alpha < 1$), the Lagrangian of the DTC can be written as
\begin{equation}
\begin{aligned}
L = &~\left( \frac{\Phi_0}{2\pi}\right)^{\!2}\left[\right.\frac{C_{33}}{2} \left(\dot{\varphi}_\mathrm{p} + \dot{\varphi}_\mathrm{m}\right)^2+\frac{C_{44}}{2} \left(\dot{\varphi}_\mathrm{p} - \dot{\varphi}_\mathrm{m}\right)^2\\
&+ 2C_{34}\dot{\varphi}_\mathrm{m}^2\left.\right]+ E_{\mathrm{J}} \cos \left(\varphi_\mathrm{p} + \varphi_\mathrm{m}\right)+E_{\mathrm{J}} \cos \left(\varphi_\mathrm{p} - \varphi_\mathrm{m}\right)\\
&+\alpha E_{\mathrm{J}}\cos\left(2\varphi_\mathrm{m}+\varphi_{\mathrm{ex}}\right)\\
= &~\left( \frac{\Phi_0}{2\pi} \right)^{\! 2}\left[C_\mathrm{c}\dot{\varphi}_\mathrm{p}^2+(C_\mathrm{c}+2C_{34}) \dot{\varphi}_\mathrm{m}^2\right]\\
& +2E_{\mathrm{J}} \cos \varphi_\mathrm{p}  \cos \varphi_\mathrm{m}+\alpha E_{\mathrm{J}}\cos\left(2\varphi_\mathrm{m}+\varphi_{\mathrm{ex}}\right), 
\end{aligned}
\end{equation}

where $\varphi_\mathrm{p} \equiv \left(\varphi_3 + \varphi_4\right)/2$ and $\varphi_\mathrm{m} \equiv \left(\varphi_3 - \varphi_4\right)/2$. 
The potential is defined as
\begin{equation}\label{Original potential}
\begin{aligned}
V = &~-2E_{\mathrm{J}} \cos \varphi_\mathrm{p}  \cos \varphi_\mathrm{m}-\alpha E_{\mathrm{J}}\cos\left(2\varphi_\mathrm{m}+\varphi_{\mathrm{ex}}\right)
\end{aligned}
\end{equation}

For a given bias flux $\varphi_{\mathrm{ex}}$, the minimum potential energy leads to
\begin{equation}\label{minimum potential}
\left\{~
\begin{aligned}
\frac{\partial V}{\partial \varphi_\mathrm{p}} =&~  2E_{\mathrm{J}} \sin \varphi_\mathrm{p} \cos \varphi_\mathrm{m} = 0, \\
\frac{\partial V}{\partial\varphi_\mathrm{m}} =&~  2E_{\mathrm{J}} \cos \varphi_\mathrm{p} \sin \varphi_\mathrm{m}+2\alpha E_{\mathrm{J}}\sin\left(2\varphi_\mathrm{m}+\varphi_{\mathrm{ex}}\right) = 0.
\end{aligned}
\right.
\end{equation}
Note that in the first equation of Eq.~(\ref{minimum potential}), assuming $\cos \varphi_\mathrm{m} = 0$ would result in $\partial^{n} V/ \partial \varphi_\mathrm{p}^n = 0$ for any order $n$. This solution does not correspond to a physically viable minimum-potential-energy position for the p-mode. Therefore, the conditions,
\begin{equation}\label{minimum potential condition}
\left\{~
\begin{aligned}
&\varphi_\mathrm{p}  = 2\pi \times k\,\, (k = \mathrm{integer}), \\
&\sin \varphi_\mathrm{m}+\alpha \sin\left(2\varphi_\mathrm{m}+\varphi_{\mathrm{ex}}\right) = 0,
\end{aligned}
\right.
\end{equation}
should be satisfied. Here, we disregard the condition $\varphi_\mathrm{p} = \pi + 2\pi \times k$, as it results in a maximum energy potential rather than a minimum.

\begin{figure}
\includegraphics[scale=1.0]{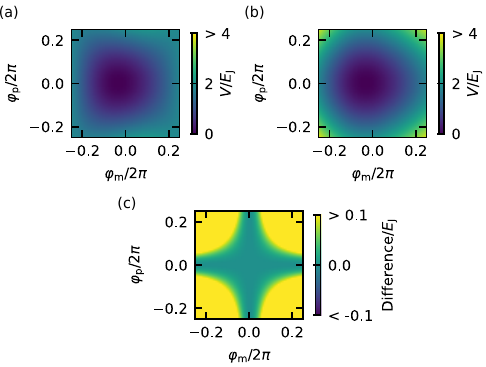}
\caption{\label{FigS11:wide} Two-dimensional potential energies $V$ at the idle bias point determined by Eq.~(\ref{idle point}): (a) original [Eq.~(\ref{Original potential})], (b) approximated [Eq.~(\ref{approx potential})], and~(c)~their difference. In (a) and (b), the potential energy value at the minimum is offset to zero for easy comparison.
}
\end{figure}

At the idle bias point, the two coupler transmons are decoupled, which is indicated by the zero coupling term, $- E_{\mathrm{J}5}\cos\left(\varphi_4-\varphi_3-\varphi_{\mathrm{ex}}\right) = 0$, at the minimum-potential-energy point. This leads to
\begin{equation}\label{idle bias condition}
\begin{aligned}
\cos\left(2\varphi_\mathrm{m}+\varphi_{\mathrm{ex}}\right) = 0.
\end{aligned}
\end{equation}
Combined with the second line of Eq.~(\ref{minimum potential condition}), the conditions,
\begin{equation}\label{minimum potential condition at idle point}
\left\{~
\begin{aligned}
& \sin \varphi_\mathrm{m}+\alpha\sin\left(2\varphi_\mathrm{m}+\varphi_{\mathrm{ex}}\right) = 0,\\
& \cos\left(2\varphi_\mathrm{m}+\varphi_{\mathrm{ex}}\right) = 0, 
\end{aligned}
\right.
\end{equation}
result in 
\begin{equation}\label{idle point}
\left\{~
\begin{aligned}
& 2\varphi_\mathrm{m}+\varphi_{\mathrm{ex}} = \pi/2 + k\pi,\\
& \sin \varphi_\mathrm{m} = \pm \alpha.
\end{aligned}
\right.
\end{equation}
In our device, $\alpha \approx  0.216$. Therefore, we estimate the bias flux at the idle bias point, $\varphi_{\mathrm{ex}}^{\mathrm{est}}/2\pi = 0.319$, which is close to the measured idle-point flux $\varphi_{\mathrm{ex}}^{\mathrm{exp}}/2\pi = 0.309$. Another solution of $\varphi_{\mathrm{ex}}^{\mathrm{est}}/2\pi = -0.319$ is also reasonable due to the symmetry of the energy potential.

Given any $\varphi_{\mathrm{ex}}$, we have
\begin{equation}
\begin{aligned}
\cos \varphi_\mathrm{m} =&~ \sqrt{1 - \sin^2 \varphi_\mathrm{m}} \\
=&~ \sqrt{1 - \alpha^2 \sin^2\left(2\varphi_\mathrm{m}+\varphi_{\mathrm{ex}}\right)}\\
\geq&~ \sqrt{1 - \alpha^2 }\\
\approx&~ 0.976.
\end{aligned}
\end{equation}
It indicates that, given the small value of $\alpha$, the inductive energy of the p-mode, whose coefficient is defined by $-2E_{\mathrm{J}} \cos \varphi_\mathrm{m}$, remains largely unchanged when we sweep the external flux. Therefore, we approximate the potential energies of the p- and m-modes separately as
\begin{equation}\label{approx potential}
\begin{aligned}
V =  -2E_{\mathrm{J}} \cos \varphi_\mathrm{p} -2E_{\mathrm{J}}\cos \varphi_\mathrm{m}-\alpha E_{\mathrm{J}}\cos\left(2\varphi_\mathrm{m}+\varphi_{\mathrm{ex}}\right).
\end{aligned}
\end{equation}

In this approximation, the term $\cos \varphi_\mathrm{p}\cos \varphi_\mathrm{m}$ is approximated as
\begin{equation}\label{approx cos}
\begin{aligned}
\cos \varphi_\mathrm{p}\cos \varphi_\mathrm{m} &\approx \left(1 - \frac{\varphi_\mathrm{p}^2}{2}+\frac{\varphi_\mathrm{p}^4}{24}\right)\left(1 - \frac{\varphi_\mathrm{m}^2}{2}+\frac{\varphi_\mathrm{m}^4}{24}\right)\\
&\approx 1 - \frac{\varphi_\mathrm{p}^2}{2}+\frac{\varphi_\mathrm{p}^4}{24} - \frac{\varphi_\mathrm{m}^2}{2}+\frac{\varphi_\mathrm{m}^4}{24} + \frac{\varphi_\mathrm{p}^2\varphi_\mathrm{m}^2}{4}\\
& \approx \cos \varphi_\mathrm{p}  + \cos \varphi_\mathrm{m} - 1 + \frac{\varphi_\mathrm{p}^2\varphi_\mathrm{m}^2}{4}.
\end{aligned}
\end{equation}
The constant value of unity and the coupling term $\varphi_\mathrm{p}^2\varphi_\mathrm{m}^2$  are omitted to obtain the potential energy described in Eq.~(\ref{approx potential}). For comparison, we plot the two-dimensional potential energies of Eq.~(\ref{Original potential}) [Fig.~\ref{FigS11:wide}(a)] and Eq.~(\ref{approx potential}) [Fig.~\ref{FigS11:wide}(b)]. Their difference [Fig.~\ref{FigS11:wide}(c)] is considerably small around the potential minimum, i.e., the point used for the circuit quantization. Therefore, this approximation is valid with a negligible coupling between the two coupler modes. While it may not yield precise results like in numerical diagonalization, it offers an intuitive understanding of the DTC scheme and fits well the energy spectra shown in Fig.~\ref{Fig1:wide}(d).

\subsection{Hamiltonian of the toy model}
We confine the following discussion to the symmetric qubit and DTC design, for simplicity, under the additional assumptions of $C_{11} = C_{22} = C_\mathrm{q}$ and $C_{13} = C_{24} = C_g$. The Lagrangian of two qubits coupled via a DTC is described as
\begin{equation}
\begin{aligned}
L = &~ K-V, \\
K  =&~\left( \frac{\Phi_0}{2\pi} \right)^{\! 2}\left[\right.\sum_{i=1}^2 \frac{C_{i i}}{2} \dot{\varphi}_i^2+\frac{C_{33}}{2} \left(\dot{\varphi}_\mathrm{p} + \dot{\varphi}_\mathrm{m}\right)^2\\
&+\frac{C_{44}}{2} \left(\dot{\varphi}_\mathrm{p} - \dot{\varphi}_\mathrm{m}\right)^2
+2C_{34}\dot{\varphi}_\mathrm{m}^2\\
&+\frac{C_{13}}{2}\left(\dot{\varphi}_1-\left(\dot{\varphi}_\mathrm{p} + \dot{\varphi}_\mathrm{m}\right)\right)^2+\frac{C_{24}}{2}\left(\dot{\varphi}_2-\left(\dot{\varphi}_\mathrm{p} - \dot{\varphi}_\mathrm{m}\right)\right)^2\left.\right]\\
V  =&~-\sum_{i=1}^2 E_{\mathrm{J}i} \cos \varphi_i -2E_{\mathrm{J}} \cos \varphi_\mathrm{p}-2E_{\mathrm{J}} \cos \varphi_\mathrm{m} \\
&-E_{\mathrm{J}5}\cos\left(2\varphi_\mathrm{m}+\varphi_{\mathrm{ex}}\right),
\end{aligned}
\end{equation}
where $C_{14}$ and $C_{23}$ are considered negligible, for simplicity, and the potential energy $V$ is approximated as in the decoupled DTC model. The kinetic energy can be written as
\begin{equation}
\begin{aligned}
K  =&~\left( \frac{\Phi_0}{2\pi} \right)^{\! 2}\left[\right.\frac{C_{11} + C_{13}}{2} \dot{\varphi}_1^2+\frac{C_{22} + C_{24}}{2} \dot{\varphi}_2^2\\
&+\frac{C_{33}+C_{44}+C_{13}+C_{24} }{2} \dot{\varphi}_\mathrm{p}^2\\
&+\frac{C_{33}+C_{44}+C_{13}+C_{24}+4C_{34}}{2} \dot{\varphi}_\mathrm{m}^2\\
&-C_{13}\dot{\varphi}_1\left(\dot{\varphi}_\mathrm{p} + \dot{\varphi}_\mathrm{m}\right)-C_{24}\dot{\varphi}_2\left(\dot{\varphi}_\mathrm{p} - \dot{\varphi}_\mathrm{m}\right) \\
&+ \left(C_{33}+C_{13} - C_{44}-C_{24}\right)\dot{\varphi}_\mathrm{p} \dot{\varphi}_\mathrm{m}\left.\right],\\
\end{aligned}
\end{equation}
where the last term $\left(C_{33}+C_{13} - C_{44}-C_{24}\right)\dot{\varphi}_\mathrm{p} \dot{\varphi}_\mathrm{m}$ will be neglected below as $C_{33}= C_{44}$ and $C_{13} = C_{24}$ in a symmetric DTC design. The canonical conjugate variables are defined as
\begin{equation}
\begin{aligned}
q_1 = \frac{\partial L}{\partial \dot{\varphi}_1} = &~\left( \frac{\Phi_0}{2\pi} \right)^{\! 2}\left[\right.\left(C_{11}+C_{13}\right)\dot{\varphi}_1\\
&-C_{13}\left(\dot{\varphi}_\mathrm{p}+\dot{\varphi}_\mathrm{m}\right)\left.\right],\\
q_2 = \frac{\partial L}{\partial \dot{\varphi}_2} = &~\left( \frac{\Phi_0}{2\pi} \right)^{\! 2}\left[\right.\left(C_{22}+C_{24}\right)\dot{\varphi}_2\\
&-C_{24}\left(\dot{\varphi}_\mathrm{p}-\dot{\varphi}_\mathrm{m}\right)\left.\right],\\
q_\mathrm{p} = \frac{\partial L}{\partial \dot{\varphi}_\mathrm{p}} = &~\left( \frac{\Phi_0}{2\pi} \right)^{\! 2}\left[\right.\left(2C_{33}+C_{13}+C_{24}\right) \dot{\varphi}_\mathrm{p}\\
&-\left(C_{13}\dot{\varphi}_1+C_{24}\dot{\varphi}_2\right)\left.\right],\\
q_\mathrm{m} = \frac{\partial L}{\partial \dot{\varphi}_\mathrm{m}} = &~\left( \frac{\Phi_0}{2\pi} \right)^{\! 2}\left[\right.\left(2C_{44}+C_{13}+C_{24} + 4C_{34}\right) \dot{\varphi}_\mathrm{m}\\
&-\left(C_{13}\dot{\varphi}_1-C_{24}\dot{\varphi}_2\right)\left.\right].\\
\end{aligned}
\end{equation}
By introducing $\mathbf{q} = \{ q_i \}$, $\mathbf{\bm{\varphi}} = \{ \varphi_i \}$, $i \in \{1,2,\mathrm{p}, \mathrm{m}\}$ and defining $\mathbf{q} = \mathbf{C}\bm{\dot\varphi}$, 
the inversion of the capacitance matrix $\mathbf{C}^{-1}$ is defined as
\begin{equation}
\begin{aligned}
\mathbf{C}^{-1} = &~\begin{pmatrix}
\frac{1}{\widetilde{C_\mathrm{q}}} & 0 & \frac{1}{\widetilde{C_{g\mathrm{p}}}} & \frac{1}{\widetilde{C_{g\mathrm{m}}}}\\ 0 & \frac{1}{\widetilde{C_\mathrm{q}}} & \frac{1}{\widetilde{C_{g\mathrm{p}}}} & -\frac{1}{\widetilde{C_{g\mathrm{m}}}} \\ \frac{1}{\widetilde{C_{g\mathrm{p}}}} & \frac{1}{\widetilde{C_{g\mathrm{p}}}} & \frac{1}{\widetilde{C_\mathrm{p}}} & 0 \\  \frac{1}{\widetilde{C_{g\mathrm{m}}}} & -\frac{1}{\widetilde{C_{g\mathrm{m}}}} & 0 & \frac{1}{\widetilde{C_\mathrm{m}}}
\end{pmatrix},
\end{aligned}
\end{equation}
where the symmetric DTC circuit parameters are assumed. Here we set the coupling terms between the two data qubits to be zero, i.e., $\mathbf{C}^{-1}_{12} = \mathbf{C}^{-1}_{21} = 0$, as they are negligible in the symmetric assumption. The differences between $\widetilde{C_{g\mathrm{p}}}$ and $\widetilde{C_{g\mathrm{m}}}$, $\widetilde{C_\mathrm{p}}$ and $\widetilde{C_\mathrm{m}}$ comes from the capacitance $C_{34}$, which is usually small with a negligible value of $C_{34}$. Consequently, the Hamiltonian is approximated as
\begin{equation}\label{approximation H}
\begin{aligned}
H  = &~\mathbf{q}^{\mathrm{T}}\bm{\dot\varphi} -L\\
= &~\frac{1}{2\widetilde{C_\mathrm{q}}} q_1^2-E_{\mathrm{J}1} \cos \varphi_1 \\
&+\frac{1}{2\widetilde{C_\mathrm{q}}} q_2^2-E_{\mathrm{J}2} \cos \varphi_2  \\
&+\frac{1}{2\widetilde{C_\mathrm{p}}} q_\mathrm{p}^2-2E_{\mathrm{J}} \cos \varphi_\mathrm{p}\\
&+\frac{1}{2\widetilde{C_\mathrm{m}}} q_\mathrm{m}^2-2E_{\mathrm{J}}\cos \varphi_\mathrm{m}-\alpha E_{\mathrm{J}}\cos\left(2\varphi_\mathrm{m}+\varphi_{\mathrm{ex}}\right)\\
&+\frac{1}{\widetilde{C_{g\mathrm{p}}}}\left(q_1+q_2\right) q_\mathrm{p} +\frac{1}{\widetilde{C_{g\mathrm{m}}}}\left(q_1 - q_2\right)q_\mathrm{m}, \\
\end{aligned}
\end{equation}
which indicates that the two data transmon modes, defined by $\{q_1, \varphi_1\}$ and $\{q_2, \varphi_2\}$, are coupled to the two coupler modes, p- and m-modes. Note that the p-mode is approximated to be a fixed-frequency transmon qubit~P, which is valid for small $\alpha$. On the other hand, the m-mode acts as a capacitively shunted flux qubit~(CSFQ)~M~\cite{yan2016flux}, whose frequency can be tuned with external flux $\varphi_{\mathrm{ex}}$. In the last term of the Hamiltonian [Eq.~(\ref{approximation H})], the coupling term between Q2 and the m-mode results in an effective capacitance with a negative value $-\widetilde{C_{g\mathrm{m}}}$, which is crucial in demonstrating the quantized Hamiltonian [Eq.~(\ref{quantized H}) in the main text] with a minus sign in front of the coupling term $g_{2\mathrm{m}}$. The coupling $g_{i\mathrm{p}}$ and $g_{i\mathrm{m}}$ can be calculated as
\begin{equation}\label{approximation g}
\begin{aligned}
g_{1\mathrm{p}} = &~\sqrt{\omega_1 \omega_\mathrm{p}}\, \frac{C_{g}}{2 \sqrt{\left(C_{\mathrm{q}}+C_{g}\right)\left(C_{\mathrm{c}}+C_{g}\right)}},\\
g_{2\mathrm{p}} = &~\sqrt{\omega_2 \omega_\mathrm{p}}\, \frac{C_{g}}{2 \sqrt{\left(C_{\mathrm{q}}+C_{g}\right)\left(C_{\mathrm{c}}+C_{g}\right)}},\\
g_{1\mathrm{m}} = &~\sqrt{\omega_1 \omega_\mathrm{m}}\, \frac{C_{g}}{2 \sqrt{\left(C_{\mathrm{q}}+C_{g}\right)\left(C_{\mathrm{c}}+2C_{34}+C_{g}\right)}},\\
g_{2\mathrm{m}} = &~\sqrt{\omega_2 \omega_\mathrm{m}}\, \frac{C_{g}}{2 \sqrt{\left(C_{\mathrm{q}}+C_{g}\right)\left(C_{\mathrm{c}}+2C_{34}+C_{g}\right)}}.\\
\end{aligned}
\end{equation}
The difference between the $g_{i\mathrm{p}}$ and $g_{i\mathrm{m}}$ can be negligible when $C_{\mathrm{c}}\gg 2C_{34}$ and $\omega_\mathrm{p} = \omega_\mathrm{m}$.

In the toy model, where the coupling mediated by the DTC is represented by a transmon-like mode P and a CSFQ-like mode M, the effective coupling strength is described by Eq.~(\ref{effctive g}). The condition $g_{\mathrm{eff}} = 0$ at the idle bias point is satisfied when $\omega_{\mathrm{p}} \approx \omega_{\mathrm{m}}$. Note that the terms for the p- and m-modes in the Hamiltonian~[Eq.~(\ref{approximation H})] become equivalent and give $\omega_{\mathrm{p}} \approx \omega_{\mathrm{m}}$ when Eq.~(\ref{idle bias condition}) holds, given $\widetilde{C_\mathrm{p}} \approx \widetilde{C_\mathrm{m}}$ due to the symmetric-circuit assumption in the toy model. Therefore, the approximate condition for the idle bias point is equivalently determined by Eq.~(\ref{effctive g}) and Eq.~(\ref{idle bias condition}).

We have derived an analytical toy model with symmetric parameters in the current design of DTC. Variation of these parameters can occur in real devices due to the fabrication error. The asymmetry of those parameters and undesired stray capacitances can result in additional coupling between the four modes, i.e., the two qubits and two coupler modes, which can only be accurately determined through numerical simulations.

\section{Qubit coherence}\label{Qubit coherence}

\begin{figure*}
\includegraphics[scale=1.05]{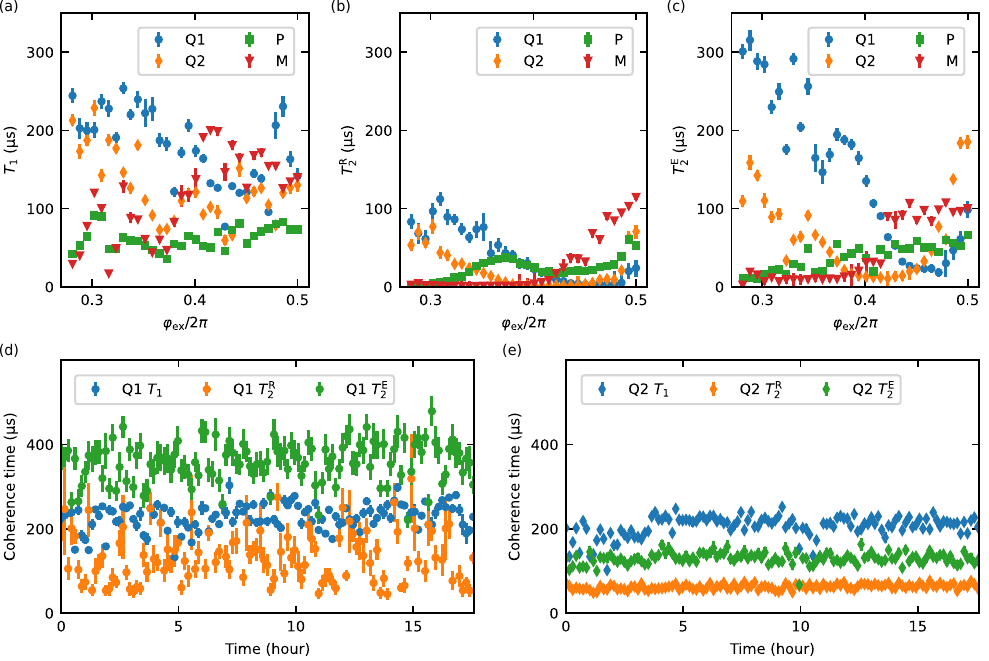}
\caption{\label{FigS2:wide} Coherence times: (a) $T_1$, (b) $T_2^\mathrm{R}$ and (c) $T_2^\mathrm{E}$ of the four-lowest excited states involving the two qubits, Q1 and Q2, and two coupler modes, P and M, as a function of the flux bias through the coupler loop. Stability of the coherence times for (d) Q1 and (e) Q2 biased at the idle bias point for 17.5 hours.} 
\end{figure*}

\begin{table}
\caption{Dressed qubit coherence at the idle bias point.}
    \centering
    \begin{tabular}{ccccc}
        \hline\hline \text { Coherence} (\textmu\text{s}) & & $T_1$ & $T_2^\mathrm{R}$ & $T_2^\mathrm{E}$  \\
        \hline \text {Q1} & & 228.6 $\pm$ 30.4 & 132.4 $\pm$ 59.9 & 358.9$\pm$ 47 \\
        \text {Q2} & & 205.3 $\pm$ 26.0 & 62.4 $\pm$ 6.1 & 129.8 $\pm$ 13.3 \\
        \hline\hline
    \end{tabular}
    \label{tab:qubit coherence}
\end{table}

By tuning the flux generated with the superconducting coil, the coherence time, including $T_1$, $T_2^\mathrm{R}$ and $T_2^\mathrm{E}$, were carefully measured for each of the four lowest excitation modes involving Q1, Q2, P and M [Figs.~\ref{FigS2:wide}(a)--(c)]. As a function of the flux through the DTC's loop, the modes are hybridized. For simplicity, however, in Fig.~\ref{FigS2:wide} we keep the labels of the dominant modes at the idle bias point. The colors in the plot corresponds to the spectra in Fig.~\ref{Fig1:wide}(d). The signal for $T_1$ measurement is fitted with an exponential decay: $P_{|1\rangle} = A + Be^{-t/T_1}$. While for the signal decay envelop of $T_2^\mathrm{R}$ and $T_2^\mathrm{E}$ measurement, a fitting function of $P_{|1\rangle} = A + Be^{-(t/T_2)^n}$ is used with $n$ as a fitting parameters. Near the idle bias point~($0.275 < \varphi_\mathrm{ex}/2\pi < 0.325$), $T_1$ of the coupler modes are lower than that of the qubits. This is due to the fact that the coupler modes have higher frequencies and more sensitive to the dielectric loss because of the geometry~\cite{gambetta2016investigating}. In the strong-coupling regime~($0.325 < \varphi_\mathrm{ex}/2\pi < 0.5$), the states of the four transmons become more hybridized, resulting in a decrease in $T_1$ for the qubits and an increase in $T_1$ for the coupler modes. The variation of both $T_2^\mathrm{R}$ and $T_2^\mathrm{E}$ mainly depends on the sensitivity of each dressed qubit to flux noise.

At the idle bias point, the stability of the coherence time were measured for qubits [Figs.~\ref{FigS2:wide}(d) and~(e)], and the averaged values are summarized in Table~\ref{tab:qubit coherence}. 
Both Q1 and Q2 demonstrate $T_1 > 200$ \textmu s, a testament to the favorable impact of fabrication and geometrical design in mitigating dielectric noise. The energy level of Q1, which is far detuned from the coupler modes, is only marginally repelled by the coupler modes at the idle bias point. As a result, the $T_2^\mathrm{R}$ and $T_2^\mathrm{E}$ of Q1 are longer than the corresponding coherence time of Q2, owing to Q1's lower sensitivity to flux noise at the idle bias point. Possibly attributed to the thermal photon fluctuations in the resonator that dominate the low-frequency noise, Q1 exhibits $T_2^\mathrm{E}/T_1 \approx 1.57$, slightly less than 2~\cite{yan2016flux}. In the mean time, for Q2, although not significantly, both $T_2^\mathrm{R}$ and $T_2^\mathrm{E}$ are further decreased due to the larger sensitivity to the flux noise, stemming from the non-negligible repulsion caused by the coupler modes.

\section{State discrimination}\label{State discrimination}

\begin{figure*}
\includegraphics[scale=1.0]{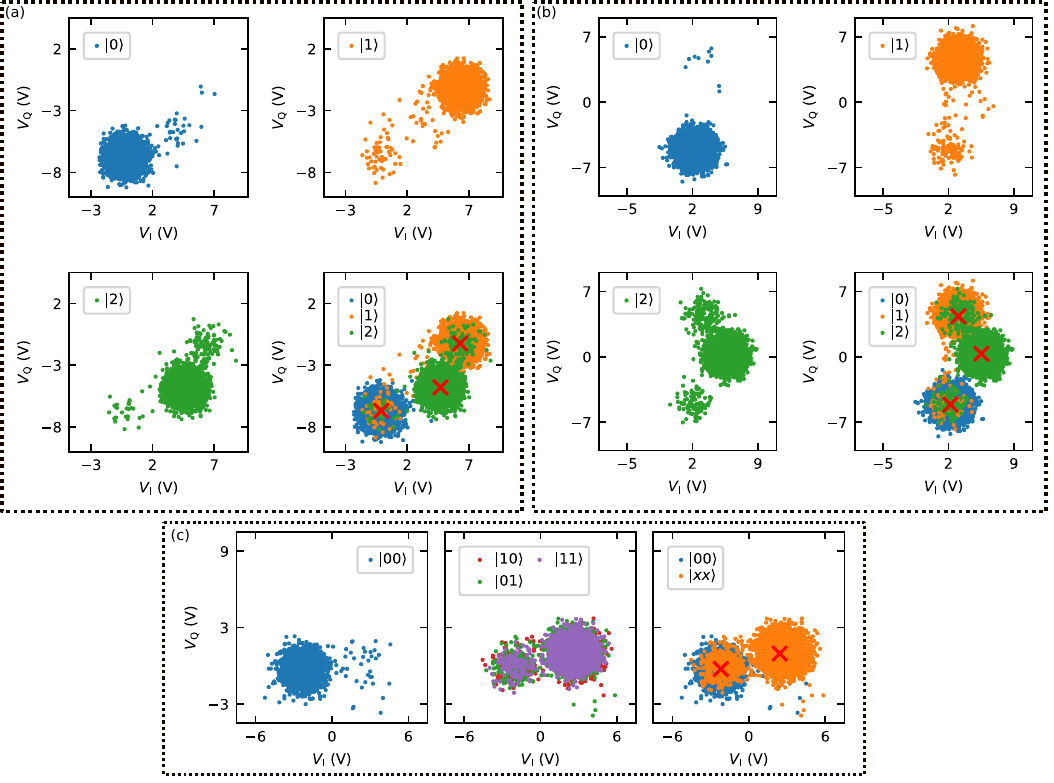}
\caption{\label{FigS3:wide} Single-shot readouts of the three resonators for the state discrimination of (a) Q1, (b) Q2 and (c) coupler modes P and M. Each point in the I--Q plane represents one demodulation result obtained from the SHFQC device. The cross markers represent the mean value for the corresponding states. In (a) and (b), each input state in the legend was prepared and measured 5000 times, while in (c) 3000 times.} 
\end{figure*}

\begin{table}
\caption{Qubit assignment probability matrix. $P(m|n)$ represents the probability of reading out the $|m\rangle$ state given the prepared state $|n\rangle$. $|xx\rangle$ includes all the excited states of the coupler modes $|\mathrm{P},\mathrm{M}\rangle$.}
\begin{tabular}{lllll}
\hline \hline & & \multicolumn{3}{c}{ Prepared state $n$} \\
\hline Q1 $P(m|n)$ & & \multicolumn{1}{c}{$|0\rangle$} & \multicolumn{1}{c}{$|1\rangle$} & \multicolumn{1}{c}{$|2\rangle$} \\
Readout state $m$ & $|0\rangle$ & 0.9933 & 0.0006 & 0.0061 \\
& $|1\rangle$ & 0.0121 & 0.9787 & 0.0092 \\
& $|2\rangle$ & 0.0060 & 0.0257 & 0.9683 \\
\hline
Q2 $P(m|n)$ & & $|0\rangle$ & $|1\rangle$ & $|2\rangle$ \\
Readout state $m$ & $|0\rangle$ & 0.9973 & 0.0017 & 0.0010 \\
& $|1\rangle$ & 0.0231 & 0.9704 & 0.0065 \\
& $|2\rangle$ & 0.0196 & 0.0398 & 0.9406 \\
\hline
Couplers $P(m|n)$ & & $|00\rangle$ & $|xx\rangle$  \\
Readout state $m$ & $|00\rangle$ & 0.9871 & 0.0129  \\
& $|xx\rangle$ &  0.0622 & 0.9378 \\
\hline \hline
\end{tabular}
\label{tab:qubit readout fidelity}
\end{table}


Accurate state discrimination with high fidelity is crucial for quantum error correction processes, notably in applications like the readout of the ancillary qubits in surface-code implementations~\cite{fowler2012surface}. Though the SPAM error associated with readout fidelity does not impede our characterization of the CZ-gate fidelity, an enhanced readout fidelity is beneficial for improving the experimental efficiency. In this work, the single-shot readout is performed at the idle bias point, where two qubits are nearly decoupled and far-detuned from the coupler modes. Therefore, we denote the qubit state with $|\mathrm{Q}i\rangle$ as it approximates the diabatic (bare) state, which is the eigenstate of the fully decoupled system~\cite{PhysRevX.11.021058}. While the two coupler modes, the p- and m-modes, are also approximated to be decoupled at the idle bias point, we denote their states with $|\mathrm{P},\mathrm{M}\rangle$ for simplicity.

We independently prepared the qubit states of $|0\rangle$, $|1\rangle$, and $|2\rangle$, followed by single-shot measurements of the corresponding resonators for Q1 and Q2 [Figs.~\ref{FigS3:wide}(a) and~(b)]. In the case of the coupler modes~P and~M, only the first excited state is taken into account, assuming that the transition to the second excited state can be deemed negligible during both single-qubit and CZ-gate implementations. Due to the shared resonator and unoptimized resonator parameters, only the ground states $|00\rangle$ can be distinguished from the other excited states $|xx\rangle$ ($|10\rangle$, $|01\rangle$ and $|11\rangle$) [Fig.~\ref{FigS3:wide}(c)]. We note that this is sufficient for leakage-error characterization, given that the computational subspace ($\{|\widetilde{0000}\rangle, |\widetilde{1000}\rangle, |\widetilde{0100}\rangle, |\widetilde{1100}\rangle\}$) comprises only the ground state of the coupler modes.

Subsequently, unsupervised learning through $k$-means clustering is employed for the states discrimination  of the qubits and coupler modes. After training the $k$-means with the single-shot results, we calculated the assignment probability matrix (Table~\ref{tab:qubit readout fidelity}). The lower fidelity associated with higher energy levels for all qubits is attributed to incoherent errors. It is worth mentioning that a higher fidelity could potentially be achieved with a faster readout by meticulously designing the resonators and optimizing the readout pulse shape~\cite{PhysRevApplied.17.044016}.

\section{Single-qubit gate fidelities}\label{Single-qubit gate fidelities}

\begin{figure}
\includegraphics[scale=1.0]{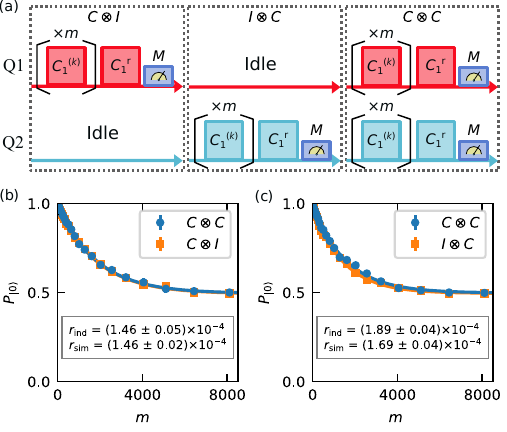}
\caption{\label{FigS4:wide} (a) Randomized-benchmarking sequence for single-qubit gates for Q1 (left), Q2 (middle), and simultaneous ones (right). $C_1^{(k)}$~($k=1,\ldots, m$) represents the randomly selected Clifford gates for a single qubit, applied $m$ times in one sequence. $C_1^\mathrm{r}$ is the recovery gate that renders the entire sequence equivalent to an identity gate. The measurement ($M$) reads out the $|0\rangle$ state with the probability $P_{|0\rangle}$, which serves as the sequence fidelity. (b) (c)~Results of the individual and simultaneous randomized benchmarking for Q1 and Q2, respectively. The data points are fitted with exponential curves, with the error bar representing the standard deviation of the sequence fidelities measured from 10 randomly selected sequences shown in~(a). Errors $r_\mathrm{ind}$ and $r_\mathrm{sim}$ measured individually and simultaneously, presented in (b) and (c), respectively, are the single-qubit gate errors calculated from the fitting results.}
\end{figure}

At the idle bias point, the residual \textit{ZZ} interaction persisted at $2\pi$$\times$$(-6.3)$~kHz. To verify that it does not impact the current single-qubit gate fidelities, we compared the fidelities measured using randomized benchmarking (RB) for both individual and simultaneous executions (Fig.~\ref{FigS4:wide}). In the individual case, we conducted single-qubit randomized benchmarking for Q1 (Q2) while maintaining Q2 (Q1) in its ground state. Subsequently, we performed simultaneous randomized benchmarking to unveil the additional error resulting from the residual \textit{ZZ} interaction. Throughout the benchmarking process, the single-qubit gates, with a duration of 48 ns for both Q1 and Q2, are repeatedly calibrated using the DRAG method. This is crucial to mitigate control errors that may arise from the device instability, particularly due to the temperature drift. 

We fitted the sequence fidelity [Figs.~\ref{FigS4:wide}(b) and~(c)] with the exponential equation
\begin{equation}
    P_{|0\rangle} = Ap^m +B,
\end{equation}
where $m$ denotes the number of randomly selected Clifford gates. The average single Clifford gate error is calculated as
\begin{equation}
    r = (1-p)(1-1/d),
\end{equation}
where $d = 2$ is the dimension of the Hilbert space for single qubit measurement. Given the average number (1.875) of physical gates per one Clifford gate calculated from its decomposition~\cite{PhysRevLett.106.180504, barends2014superconducting}, the single-qubit gate error $r_{\text{SQ}}$ equals to $r/1.875$. The high single-qubit gate fidelities for both Q1 and Q2 can be attributed to their high coherence at the idle bias point. Comparing the individual and simultaneous gate fidelities for either Q1 or Q2, the nearly identical gate errors suggests that the \textit{ZZ} interaction of $2\pi$$\times$$(-6.3)$~kHz is not the current limiting factor for single-qubit gate fidelity. The even higher fidelity for Q2 in the simultaneous measurement compared to the individual case could be attributed to variation of the coherence time.

With the coherence measurement at the idle bias point (see Appendix~\ref{Qubit coherence}), the estimation for the single-qubit-gate incoherent error ($r_\mathrm{SQ}^\mathrm{incoherent}$) is given by~\cite{PhysRevX.13.031035, PhysRevApplied.3.044009}
\begin{equation}
    r_\mathrm{SQ}^\mathrm{incoherent} = \frac{t_\text{gate}}{3}\left(\frac{1}{T_1} + \frac{1}{T_\phi}\right),
\end{equation}
where $t_\text{gate} = 48$ ns represents the single-qubit-gate duration used in our case. $T_1$ is directly measured while $T_\phi$ is estimated as $1/T_\phi=1/T_2^\mathrm{E}-1/2T_1$. The influence originating from flux noise is considered negligible, given the long $T_2^\mathrm{R}$. This is attributed to the fact that the two qubits are less sensitive to flux variations in the vicinity of the system idle bias point. The errors introduced by $T_1$ and $T_\phi$ for Q1 are $7.0\times 10^{-5}$ and $6\times10^{-6}$, respectively. Therefore, the incoherent error contributes only $\sim$50\% to the total gate error. We posit that the residual error arises from coherent errors attributed to the instability of the control pulse. This rationale underscores our repeated implementation of single-qubit gate calibration during both single- and two-qubit gate randomized benchmarking. For Q2, the errors introduced by $T_1$ and $T_\phi$ are $7.8\times 10^{-5}$ and $8.3\times10^{-5}$, respectively. While the total incoherent error constitutes more than $85\%$ of the total gate error, we believe that the larger error induced by $T_\phi$ may be overestimated due to its estimation influenced by the flux noise.

\section{\textit{Z}-pulse distortion calibration}\label{$Z$-pulse distortion calibration}

\begin{figure}
\includegraphics[scale=1.00]{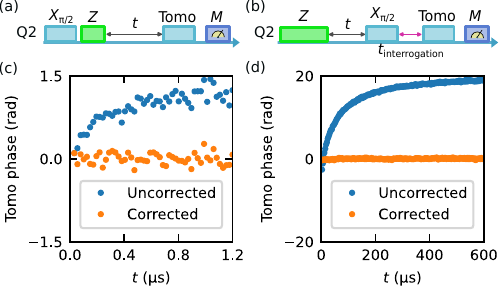}
\caption{\label{FigS5:wide} (a) Sequence for measuring the short-term distortion of a \textit{Z}-pulse. A square-shaped \textit{Z}-pulse of 48 ns is applied between a Ramsey-like sequence with varying the duration $t$ after the \textit{Z}-pulse. ``Tomo'' stands for a single-qubit gate for quantum phase tomography. (b) Sequence for measuring the long-term distortion. A square-shaped \textit{Z}-pulse of 10 \textmu s is first applied on Q2. Following the variation of a duration $t$, a Ramsey-like sequence with a fixed interrogation time ($\sim$500~ns) and phase tomography is employed to discern the enduring impact of the \textit{Z}-pulse.  (c) (d) Extracted phase accumulations due to distortion with correction (orange circles) and without correction (blue circles) for (c) short- and (d) long-term cases, respectively.
}
\end{figure}

\begin{table}
    \caption{Flux-transient parameters for pulse distortion calibrated either by short-term or long-term measurement.}
    \centering
    \begin{tabular}{lllll}
    \hline\hline \multicolumn{2}{c}{\text{Short-term calibrated}} & & \multicolumn{2}{c}{\text{Long-term calibrated}}\\
    \cline{ 1-2 }\cline { 4-5 }\multicolumn{1}{c}{$\tau$ (ns)} & \multicolumn{1}{c}{\text{Amplitude}} & & \multicolumn{1}{c}{$\tau$ (\textmu s)} & \multicolumn{1}{c}{\text{Amplitude}}\\
    \hline 603.3 &~~$-0.0104$ & & 400.5 &~~0.12 \\
    79.45 &~~$-0.0137$ & & 71.02 &~~0.038 \\
     & & & 13.60 &~~0.00525 \\
    \hline \hline
    \end{tabular}
    \label{tab:pulse distortion}
\end{table}

We followed and extended the idea for correcting the \textit{Z}-pulse distortion, as detailed in Ref.~\citenum{PhysRevX.11.021058}. A perfectly square-shaped \textit{Z}-pulse ideally ceases when it is completed. However, due to distortion, its lasting influence persists even after it finished~\cite{rol2020time}. Consequently, the turn-off transients of a square-shaped \textit{Z}-pulse with a fixed amplitude and duration are measured to calibrate the flux-transient parameters for predistortion.

A tunable qubit, whose frequency is sensitive to the applied flux through the fast flux line, is typically employed as a sensor for capturing the step response. Normally, the coupler is utilized because its frequency can be directly tuned by the applied \textit{Z}-pulse. However, in the DTC scheme, the qubit Q2 could also be employed. This is due to its frequency being significantly repelled by the coupler energy level in the range of $0.375 < \varphi_\mathrm{ex}/2\pi < 0.45$, which demonstrates an indirect susceptibility to the applied flux. This alternative choice makes it possible to perform distortion calibration without the need for the readout resonator for coupler transmons.

The transients of the \textit{Z}-pulse can be characterized using a Ramsey-type experiment [Fig.~\ref{FigS5:wide}(a)]. Typically, a $\pi/2$ pulse is applied in the beginning to create a superposition state~$\left( |0\rangle +|1\rangle \right) /\sqrt{2}$ of Q2. The turn-off transients of the \textit{Z}-pulse will accumulate a relative phase on the state of Q2, $\left(|0\rangle  + e^{i\phi(t)} |1\rangle \right)/\sqrt{2}$, where the $\phi(t)$ is finally measured by state tomography. With the step response described by multiple exponential combination, the $\phi(t)$ can be expressed as
\begin{equation}
    \phi(t)=\phi_0 \sum_k a_k\left(e^{-\left(t / \tau_k\right)} - e^{\left.-\left(t+\tau_{\text {pulse }}\right) / \tau_k\right)}\right),
\end{equation}
where $\phi_0$ is a $\text{constant}$ and $\tau_{\text {pulse }}$ is the duration of the \textit{Z}-pulse. The parameters $a_k$ and $\tau_k$ are the flux-transient parameters fitted from the measured phase and used for predistortion.

\begin{figure}
\includegraphics[scale=1.0]{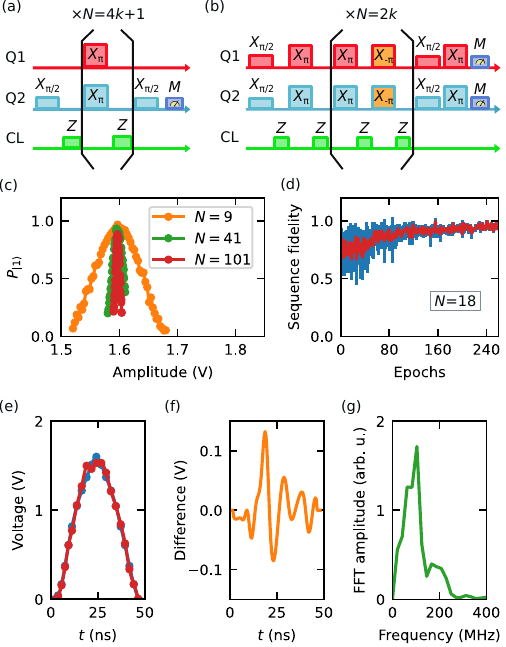}
\caption{\label{FigS6:wide} (a) JAZZ-$N$ pulse sequence for the amplitude calibration of the \textit{Z}-pulse in a Slepian shape, where the repetition $N = 4k + 1$ with integer $k$. The fidelity of the sequence is assessed based on the population of Q2, $P_{|1\rangle}$. (b) JAZZ2-$N$ pulse sequence for the \textit{Z}-pulse shape calibration and optimization, where the repetition $N = 2k$ with integer $k$. The fidelity of the sequence, assessed through $P_{|00\rangle}$, serves as the reward for the RL algorithm. (c) Measurement results according to the JAZZ-$N$ sequence, as depicted in (a), aiming to finely tune the amplitude of the Slepian pulse. (d) Sequence fidelity evaluated, utilizing the JAZZ2-$N$ sequence~($N=18$) illustrated in~(b), along the training epochs of the RL algorithm. The red line illustrates the average fidelities, while the blue lines represent error bars, each calculated from 20 trial shapes generated by the agents following the policy prescribed by the RL algorithm. (e) Representative instance showcasing the Slepian pulse shape (blue circles) as the initial shape for the RL algorithm, alongside the optimized pulse shape (red circles). The connecting lines illustrate the cubic spline interpolations. (f)~Difference calculated from the interpolation results depicted in (e). (g)~FFT amplitude of the shape difference depicted in (f).
}
\end{figure}

Firstly, we maintained the external flux bias at $\varphi_\mathrm{ex}/2\pi \approx 0.42$ using the superconducting coil. A pulse sequence, as illustrated in Fig.~\ref{FigS5:wide}(a), is employed to measure the short-term transients of the \textit{Z}-pulse, where the duration $t$ is constrained by qubit decoherence. Because of the larger flux sensitivity of Q2 at this point, $T_2^\mathrm{R}$ is only $\sim$3 \textmu s. Therefore, for long-term distortion, especially those exceeding the coherence time of the qubit, a sequence depicted in Fig.~\ref{FigS5:wide}(b) can be employed. To avoid passing through the anti-crossing with the applied \textit{Z}-pulse in sequence (a), a smaller amplitude of the \textit{Z}-pulse is chosen. Although the accumulated phase $\phi(t)$ is relatively small, it is sufficient for the short-term transient calibration. While in sequence (b), the amplitude of the \textit{Z}-pulse can be larger, as all the qubits are at their ground states before the end of the \textit{Z}-pulse. To increase the accumulated phase $\phi(t)$ for a better signal-to-noise ratio (SNR), the duration of the \textit{Z}-pulse is chosen to be relatively longer for long-term transient measurement. Figures~\ref{FigS5:wide}(c) and (d) displays the phases measured with or without \textit{Z}-pulse corrections for short- and long-term transients, respectively. The flux-transient parameters (Table~\ref{tab:pulse distortion}) were calibrated and utilized for the predistortion during the CZ-gate implementation.

\section{CZ-gate optimization by reinforcement-learning algorithm}\label{CZ gate optimization by RL algorithm}

To achieve an optimal CZ gate, we employ a model-free optimization process driven by the RL algorithm. In particular, we employ  the code adapted from Ref.~\citenum{quantumcontrolrl}, utilizing the well-known proximal-policy-optimization (PPO) algorithm~\cite{schulman2017proximal} as the agent. The agent is trained to provide the pulse shape of the CZ gate, autonomously learning to optimize it through feedback in the form of rewards. We deployed this agent on the same computer for the experiment without utilizing GPU, typically completing the training process in about 2 hours. We note that the primary time consumption are associated with waveform uploading into the device and policy updates in the agent.

As shown in Fig.~\ref{Fig4:wide}(a), a CZ gate is decomposed into three gates, showcasing a \textit{Z}-pulse injected into the DTC's loop to determine the CZ phase, accompanied by two VZ gates applied to the respective qubits to compensate for the single-qubit rotation. The VZ-gate phase is highly sensitive to the shape of the \textit{Z}-pulse, attributed to the significant repulsion in the energy levels of the qubits from those of the coupler modes. Therefore, a joint optimization of the CZ and VZ phases becomes extremely difficult in the DTC scheme. Instead, we divide the optimization process into two steps. Initially, we optimize the \textit{Z}-pulse shape to achieve a CZ phase of $\pi$, following the sequence advanced by the JAZZ method, without the requirement for applying the VZ gates. Subsequently, the optimization focuses on the VZ-gate phases while keeping the \textit{Z}-pulse optimized and held constant. In each optimization step, the relevant parameters are initially calibrated through a physics-based process. Following that, the calibrated parameters serve as the initial values for the subsequent RL algorithm. Upon completion, the optimized CZ gate is employed in the quantum process tomography and gate-fidelity characterization by the IRB experiment.

\subsection{CZ-phase optimization}\label{CZ phase optimization}

The JAZZ-$N$ sequence depicted in Fig.~\ref{FigS6:wide}(a) functions as a physics-based process utilized for the calibration of the \textit{Z}-pulse for a CZ gate. Adapted from the JAZZ method, a repeated ($\pi$--$Z$)$^N$ pattern is employed to enhance the sequence fidelity sensitivity to pulse amplitude. In this experiment, a Slepian shape [Fig.~\ref{FigS6:wide}(e)] is initially calculated with a desired duration and unit amplitude. Subsequently, the amplitude of the \textit{Z}-pulse is systematically adjusted through scaling in the AWG device during the experiment. Without considering the decoherence, the evolution of the states of two qubits based on the JAZZ-$N$ sequence can be interpreted as
\begin{widetext}
\begin{equation}
\begin{split}
|00\rangle \xrightarrow{IX_{\pi/2}}&~|0\rangle\frac{1}{\sqrt{2}}\left(|0\rangle -i |1\rangle\right) = \frac{1}{\sqrt{2}}\left(|00\rangle -i |01\rangle\right)\\
            \xrightarrow{Z}&~\frac{1}{\sqrt{2}}\left(e^{i\theta_{00}}|00\rangle -i e^{i\theta_{01}} |01\rangle\right)\\
            \xrightarrow{X_{\pi}X_{\pi}}&~\frac{1}{\sqrt{2}}\left( e^{i\theta_{00}}|11\rangle -i e^{i\theta_{01}} |10\rangle\right)\\
            \xrightarrow{Z}&~\frac{1}{\sqrt{2}}\left( e^{i \left(\theta_{00}+\theta_{11}\right)}|11\rangle -i e^{i\left(\theta_{01}+\theta_{10}\right)} |10\rangle\right)\\
            \xrightarrow{\text{remove an overall phase} ~~ e^{i\left(\theta_{01}+\theta_{10}\right)}} &~\frac{1}{\sqrt{2}}\left( e^{i\theta_{\mathrm{CZ}}}|11\rangle -i|10\rangle\right)\\
            \xrightarrow{\text{accumulate phase with}~ (N-1)/2 = 2k ~\text{more times}}&~\frac{1}{\sqrt{2}}|1\rangle\left( e^{i(2k+1)\theta_{\mathrm{CZ}}}|1\rangle -i|0\rangle\right)\\
            \xrightarrow{IX_{\pi/2}}&~\frac{1}{\sqrt{2}}|1\rangle\left(-i\left(1 + e^{i(2k+1)\theta_{\mathrm{CZ}}} \right)|0\rangle - \left(1 - e^{i(2k+1)\theta_{\mathrm{CZ}}} \right)|1\rangle\right)\\
            \xrightarrow{\text{Measurement of Q2}}&~ P_{|1\rangle} = \frac{1 - \cos\left((2k+1)\theta_{\mathrm{CZ}}\right)}{2}.\\
\end{split}
\end{equation}
\end{widetext}
Hence, the sequence fidelity, defined as $P_{|1\rangle}$ of Q2, is expressed as $(1 - \cos\left((2k+1)\theta_{\mathrm{CZ}}\right))/2$. Note that the single-qubit rotation is automatically omitted, thereby decoupling the fidelity from the VZ gates. Considering instances where $N = 9, 41, 101$ with corresponding values of $k = 2, 10, 25$, the measurement of $P_{|1\rangle}$ is conducted by adjusting the amplitude of the \textit{Z}-pulse, as depicted in Fig.~\ref{FigS6:wide}(c). As the repetition count $N$ increases, a more finely calibrated amplitude is obtained.

Figure~\ref{FigS6:wide}(e) displays a fine-tuned Slepian shape with a duration of 48 ns, featuring 2 ns of zero padding at both the beginning and end of the pulse. Its original shape with unit amplitude is calculated based on the method described in Ref.~\citenum{PhysRevA.90.022307}. Specifically, we start from the control parameter $\theta$ with the definition
\begin{equation}
    \theta=\arctan \left(H_x / H_z\right), 
\end{equation}
given a reduced two-state Hamiltonian,
\begin{equation}
H=H_x \sigma_x+H_z \sigma_z=\left(\begin{array}{rr}
H_z & H_x \\
H_x & -H_z
\end{array}\right).
\end{equation}
A smooth tuning of $\theta(t)$ is optimized to form a CZ gate, striking a balance between speed and adiabaticity. By selecting the Slepian function $S(t)$ to derive the optimal pulse shape, the desired evolution of $\theta (t)$ can be described as
\begin{equation}
    \frac{d\theta}{dt}=S(t).
\end{equation}
Therefore, the evolution of the $H_z/H_x = 1/\tan\theta(t)$ can be obtained with $\theta(t) = \int S(t) dt$. To relate $\theta(t)$ to the applied external flux, specifically the voltage $z(t)$ of the arbitrary waveform generated by the AWG, we approximate $H_z/H_x  \approx C z^2(t)$. The coefficient $C$ can be any constant, as we intend to normalize the final pulse shape to begin and end at zero, with a unit peak value. Despite the approximation, we observed a commendable performance of the generated Slepian shape in our experiment. Additionally, it functions solely as an initial pulse, and a superior shape will be optimized using the RL algorithm. We note that the initial pulse shape can take various forms, such as Gaussian or cosine shapes. These shapes merely introduce model bias, which does not impact the final performance of the optimized shape but influences only the efficiency of the optimization process.

To enhance the sequence sensitivity to the \textit{Z}-pulse shape and increase robustness to single-qubit gate errors, we employ the sequence, denoted as JAZZ2-$N$ sequence and depicted in Fig.~\ref{FigS6:wide}(b), to generate the sequence fidelity as the reward for the PPO agent. Similar to the previous sequence, the evolution of the two-qubit states based on JAZZ2-$N$ sequence could be derived as
\begin{widetext}
\begin{equation}
\begin{split}
|00\rangle \xrightarrow{X_{\pi/2}X_{\pi/2}} &~\frac{1}{\sqrt{2}}\left(|0\rangle -i |1\rangle\right)\frac{1}{\sqrt{2}}\left(|0\rangle -i |1\rangle\right) = \frac{1}{2}\left(|00\rangle -i |01\rangle-i |10\rangle- |11\rangle\right)\\ 
            \xrightarrow{Z}&~\frac{1}{2}\left(e^{i\theta_{00}}|00\rangle -i e^{i\theta_{01}}|01\rangle-i e^{i\theta_{10}}|10\rangle- e^{i\theta_{11}}|11\rangle\right)\\
            \xrightarrow{X_{\pi}X_{\pi}}&~\frac{1}{2}\left(e^{i\theta_{00}}|11\rangle -i e^{i\theta_{01}}|10\rangle-i e^{i\theta_{10}}|01\rangle- e^{i\theta_{11}}|00\rangle\right)\\
            \xrightarrow{Z}&~\frac{1}{2}\left(e^{i\left(\theta_{00} + \theta_{11}\right)}|11\rangle -i e^{i\left(\theta_{01} + \theta_{10}\right)}|10\rangle-i e^{i\left(\theta_{10} + \theta_{01}\right)}|01\rangle- e^{i\left(\theta_{11} + \theta_{00}\right)}|00\rangle\right)\\
            \xrightarrow{\text{remove an overall phase} ~~ e^{i\left(\theta_{01}+\theta_{10}\right)}} &~\frac{1}{2}\left(e^{i\theta_{\mathrm{CZ}}}|11\rangle -i|10\rangle-i|01\rangle- e^{i\theta_{\mathrm{CZ}}}|00\rangle\right)\\
            \xrightarrow{\text{accumulate phase with}~ N = 2k ~\text{more times}}&~\frac{1}{2}\left(e^{i(2k+1)\theta_{\mathrm{CZ}}}|11\rangle -i|10\rangle-i|01\rangle- e^{i(2k+1)\theta_{\mathrm{CZ}}}|00\rangle\right)\\
            \xrightarrow{X_{\pi/2}X_{\pi/2}}&~\frac{1}{4}\left( \right. e^{i(2k+1)\theta_{\mathrm{CZ}}}\left( -|00\rangle -i |01\rangle-i |10\rangle + |11\rangle\right) \\
            &~+ i\left( -i|00\rangle - |01\rangle + |10\rangle- i|11\rangle\right)\\
            &~+ i\left( -i|00\rangle - |10\rangle + |01\rangle- i|11\rangle\right)\\
            &~-e^{i(2k+1)\theta_{\mathrm{CZ}}}\left(|00\rangle -i |01\rangle-i |10\rangle- |11\rangle\right) \left.\right)\\
            \xrightarrow{\text{amplitude of }|00>} &~\frac{1}{4}\left( - e^{i(2k+1)\theta_{\mathrm{CZ}}} + 1 + 1 - e^{(2k+1)\theta_{\mathrm{CZ}}}\right)\\
            \xrightarrow{\text{Measurement of Q1 and Q2} }&~  P_{|00\rangle} = \frac{1 - \cos\left((2k+1)\theta_{\mathrm{CZ}}\right)}{2}.\\
\end{split}
\end{equation}
\end{widetext}
Therefore, the sequence fidelity, defined as $P_{|00\rangle}$, reaches unity when $\theta_{\mathrm{CZ}} \xrightarrow{} \pi$. Figure~\ref{FigS6:wide}(d) illustrates the increasing fidelity as the training progresses. In each epoch, the PPO agent generates 20 waveforms based on the initial pulse and its policy. Initially, when the agent is not well-trained, it generates nearly random pulse shapes with larger variations, leading to lower fidelities. After approximately 100 epochs of training, the agent produces all trial pulses that closely resemble the optimal one, resulting in high fidelity with smaller variations. At times, for a more refined optimization, we incrementally increased $N$ and carried out the optimization process using the previously optimized pulse shape.

Figure~\ref{FigS6:wide}(e) shows a representative example of a Slepian pulse along with its optimized shape. Throughout the optimization process, the PPO agent generates 20 points for each pulse. When constructing the waveform using the AWG, a sampling rate of 2 GSa/s is employed, incorporating cubic interpolations of the aforementioned 20 points. We calculated the difference between the interpolated pulse shape~[Fig.~\ref{FigS6:wide}(f)] and analyzed it in the frequency domain using fast Fourier transform (FFT)~[Fig.~\ref{FigS6:wide}(g)]. We observed that the agent is attempting to incorporate specific frequency components into the original pulse shape. This behavior implies a potential effort to compensate for the residual pulse distortion characterized by various short transient times with small amplitudes. Alternatively, this could be explained by the suppression of transitions in the high-energy levels around the anti-crossings, forming an adiabatic process.

\subsection{VZ-phase optimization}\label{VZ phase optimization}

\begin{figure}
\includegraphics[scale=1.05]{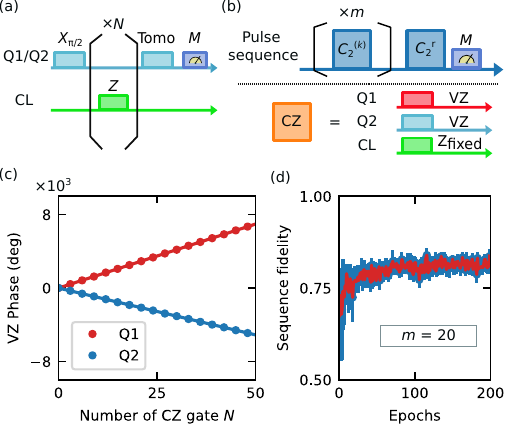}
\caption{\label{FigS7:wide} (a) Ramsey-like pulse sequence designed for the VZ-phase calibration. When either Q1 or Q2 is measured, the other is held in the ground state. (b) Pulse sequence similar to the randomized benchmarking of the two-qubit Clifford gates. The sequence fidelity, evaluated using $P_{|00\rangle}$, serves as the reward in the optimization process driven by the RL algorithm. The pulse shape of the flux applied to the DTC's loop is pre-optimized and held fixed during this measurement. (c) Separately measured single-qubit phases respectively accumulated on Q1 and Q2 in the sequence depicted in (a). The solid lines shows linear fittings of the corresponding data, which gives the phase per cycle of $139.33^\circ$ and $-102.05^\circ$ for Q1 and Q2, respectively. (d) Evaluation of the sequence fidelity, using the sequence depicted in (b), across the training epochs of the RL algorithm optimization for the VZ gates. The red line illustrates the average fidelities, while the blue lines represent error bars, each calculated from 20 trial VZ-gate pairs ([Q1, Q2]) generated by the agents.
}
\end{figure}

Once the \textit{Z}-pulse is calibrated, it is kept constant in the subsequent optimization for the two VZ gates. An initial physics-based calibration, illustrated in Fig.~\ref{FigS7:wide}(a), is employed to independently measure the \textit{Z}-pulse-induced phases ($\theta_1$ and $\theta_2$) on Q1 and Q2, respectively. The \textit{Z}-pulse induces a relative phase $\theta_{1/2}$, which is iteratively repeated $N$ times, between the superposition states created by the initial $\pi/2$ pulse. Subsequently, a state tomography is performed to extract the accumulated phase $N\theta_{1/2}$ from the qubit population measurements. Figure~\ref{FigS7:wide}(c) displays typical measurement results, with a linear fitting applied to extract the value of $\theta_{1/2}$. We emphasize that an imprecise calibration of the qubit frequency introduces an additional phase, impeding the accurate calibration of the VZ gate. This issue can be addressed either by employing an echo-type sequence ($\pi$/2--$Z$--$\pi$--$I$--Tomo) or through the calibration of the qubit frequency in advance. Besides, a slight imprecision in the VZ gates are acceptable based on the current characterization, as the optimization process will be conducted to fine-tune and enhance its accuracy.

Utilizing the optimized \textit{Z}-pulse and initially calibrated VZ gates, we evaluate the sequence fidelity from the two-qubit randomized benchmarking [Fig.~\ref{FigS7:wide}(b)] to further optimize the VZ gates. The \textit{Z}-pulse remains fixed during the optimization of the VZ gates, with the initially calibrated VZ gates serving as the initial values for the PPO agent. A typical optimized process is illustrated in Fig.~\ref{FigS7:wide}(d), showcasing an increase in sequence fidelity as the agent undergoes its training. During each training epoch, the agent generates 20 pairs of VZ gates for Q1 and Q2. Subsequently, a set of 5 randomly selected Clifford-gate sequences, with $m = 20$, is employed for measurements. The average fidelity of the set is then obtained alongside each pair of the aforementioned VZ gates. As a result, the sequence fidelity exhibits a larger variation at the beginning of the training and then gradually stabilizes with smaller variations after 100 epochs. A larger value for $m$ could be employed to finely tune the VZ gate. However, we observe that its accuracy is finally constrained by the readout fidelity. It is worth noting that further optimization of VZ gates through randomized benchmarking is crucial for minimizing the phase error and enhancing the CZ-gate fidelity.


In conclusion of this section, we independently optimize the \textit{Z}-pulse and VZ phases using the RL algorithm. The optimized CZ gate not only results in an adiabatic process that reduces the leakage error but also diminishes the phase error. In addition to achieving high CZ-gate fidelity, it is crucial to emphasize training efficiency, especially in the context of scalability. Additional efforts should be devoted to a more rational design of the PPO agent to expedite convergence. Furthermore, an advanced setup with GPU acceleration is desirable to speed up calculations.

\section{Quantum process tomography}\label{Quantum process tomography}

\begin{figure*}
\includegraphics[scale=1.0]{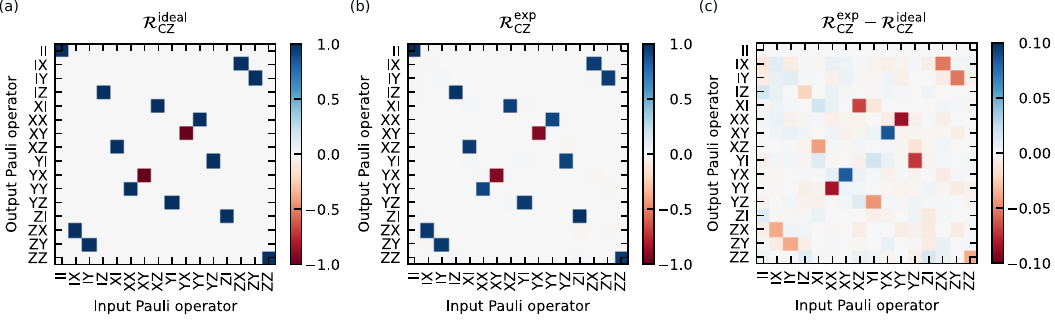}
\caption{\label{FigS8:wide} Quantum process tomography. (a) Ideal CZ-gate PTM $\cal{R}_\mathrm{CZ}^\mathrm{ideal}$. (b) Reconstructed CZ-gate PTM $\cal{R}_\mathrm{CZ}^\mathrm{exp}$. (c) Difference between the reconstructed and ideal CZ-gate PTMs, $\cal{R}_\mathrm{CZ}^\mathrm{exp}- \cal{R}_\mathrm{CZ}^\mathrm{ideal}$.
}
\end{figure*}

Quantum process tomography (QPT) is employed to assess the performance of a quantum gate. However, the fidelity estimated from QPT tends to be underestimated due to the presence of SPAM errors, especially given the current high gate performance. While a pure gate fidelity can be measured through the randomized benchmarking experiment, QPT demonstrates its advantage in constructing a more intuitive matrix representation of a quantum operator. We employ a Pauli transfer matrix (PTM) $\mathcal{R}$~\cite{greenbaum2015introduction} to represent the quantum process of our optimized CZ gate. The PTM contains only real numbers within the range $[-1, 1]$ as its matrix elements, allowing for an intuitive representation as a 2D map. In addition, the trace preservation of a quantum map can be succinctly demonstrated as $(\mathcal{R})_{0j} = \delta_{0j}$, representing physical process that do not cause information leakage.

By using the QPT, a PTM $\cal{R}_\mathrm{CZ}$ of the CZ gate can be compiled over an over-complete measurement set~\cite{PhysRevLett.109.060501}. The PTM of CZ gate can be represented as $(\mathcal{R}_\mathrm{CZ})_{ij} = \langle i |\mathcal{R}_\mathrm{CZ}| j\rangle$,  where $i \in \{0, 1, 2, ..., 15\}$ corresponding to the 16 Pauli basis $\{I, X, Y, Z\}^{\otimes2}$. Given $\Vec{p}$ represents the vectorized state density matrix in the Pauli basis, the state evolution with a CZ gate applied can be described as $\Vec{p}_\text{out} = \mathcal{R}_\mathrm{CZ}\,\Vec{p}_\text{in}$. Therefore, a series of state-tomography experiments can be performed to reconstruct the matrix elements of $\mathcal{R}_\mathrm{CZ}$. Based on this idea, initial states projected on axes in $\{X+, X-, Y+, Y-, Z+, Z-\}^{\otimes2}$ are independently prepared by the gates in $\{Y_{\pi/2}, -Y_{\pi/2}, -X_{\pi/2}, X_{\pi/2}, I, X_\pi\}^{\otimes2}$, resulting in a total of 36 initial states. After applying the CZ gate to each initial state, a projection measurement is performed on the $\{X+, X-, Y+, Y-, Z+, Z-\}^{\otimes2}$ axes, resulting in a total of $36 \times 36$ measurement outcomes. For each initial state, a final-state density matrix is reconstructed through state tomography, utilizing the projection measurement results on the $\{X+, X-, Y+, Y-, Z+, Z-\}^{\otimes2}$ axes. Subsequently, the density matrix is vectorized in the Pauli basis. The total of 36 final-state vectors, along with their initial states, are then employed to reconstruct the PTM using a maximum-likelihood algorithm~\cite{PhysRevA.98.062336, forestbenchmarking}.

Figure~\ref{FigS8:wide} shows a representative QPT result of our two-qubit gate. The close resemblance between the ideal and experimental PTM intuitively indicates that the processing gate is a CZ gate. The gate fidelity 0.959 calculated based on the QPT results is given by~\cite{PhysRevLett.109.060501}
\begin{equation}
\bar{F}=\frac{\operatorname{Tr}\left[\mathcal{R}_{\text {ideal }}^\mathrm{T} \mathcal{R}\right] +d}{d(d+1)}, 
\end{equation}
where $d = 4$ for the two-qubit system. The difference between the ideal PTM and the experimentally reconstructed one reveals that $\left(\mathcal{R}_\mathrm{CZ}^\mathrm{exp}\right)_{0j}=\delta_{0j}$, corresponding to the trace-preserving assumption utilized in the maximum-likelihood algorithm. Although a leakage error exists in the CZ gate, its comparatively smaller magnitude compared to the SPAM error, is believed not to significantly alter the conclusions drawn from the QPT experiment.

\section{\Add{CZ-gate fidelity}}\label{CZ gate evaluation}

It is known that the widely used randomized benchmarking~(RB) protocol needs to be modified 
in the presence of leakage errors~\cite{PhysRevA.97.032306}. Here we explain the leakage RB~(LRB) and the protocol used in this work.

\begin{figure}
	\includegraphics[width=7cm]{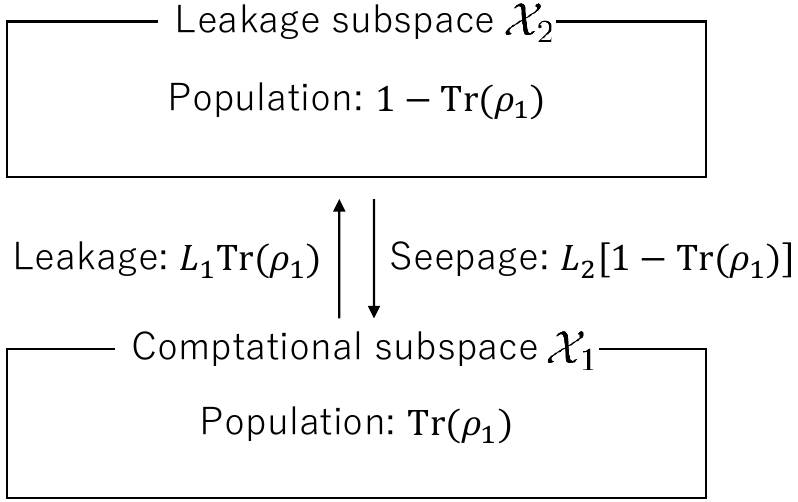}
	\caption{Leakage error model. }
	\label{fig-LRB}
\end{figure}

The system with leakage errors is modeled using 
the computational subspace $\mathcal{X}_1$ 
(the two-qubit subspace in the present case) 
and the leakage subspace $\mathcal{X}_2$ 
(the total state space is $\mathcal{X}_1 \oplus \mathcal{X}_2$), 
as shown in Fig.~\ref{fig-LRB}, where 
$\rho_1$ denotes the $\mathcal{X}_1$ submatrix of the density matrix.
In addition to errors in the computational subspace $\mathcal{X}_1$, 
there are two kinds of errors: leakage errors 
(population transfer from $\mathcal{X}_1$ to $\mathcal{X}_2$) with rate $L_1$ and 
seepage errors (population transfer from $\mathcal{X}_2$ to $\mathcal{X}_1$) 
with rate $L_2$.

Errors per Clifford gate in LRB are modeled by the following evolution of the density matrix as a function of the number of randomly selected Clifford gates, $m$: 
\begin{align}
\rho_{1\,} \! (m+1)
=&
\left( 1-L_1 \right) \,\! \mathcal{E}_D[\rho_{1\,} \! (m)]
\nonumber \\
& + L_2 \! \left\{ 1 - \mathrm{Tr}[\rho_{1\,} \! (m)] \right\} \frac{I_1}{d},
\label{eq-error1}
\end{align}
where 
$I_1$ and ${d=4}$ are the identity matrix and dimension of $\mathcal{X}_1$, respectively. $I_1/d$ describes the maximally mixed state for $\mathcal{X}_1$, and
\begin{align}
\mathcal{E}_D (\rho_1)
=
\left( 1-p_D \right) \! \rho_1 
+ p_D \mathrm{Tr}(\rho_1) \frac{I_1}{d}
\label{eq-error2}
\end{align}
describes the trace-preserving depolarizing error in $\mathcal{X}_1$ 
with probability $p_D$. 
The isotropy of the error model comes from twirling over Clifford gates.

In LRB, we consider the following two survival probabilities: 
the probability of the ideal output state $|\psi_{\mathrm{id}} \rangle$, 
${P_{\mathrm{id\,}}=\langle \psi_{\mathrm{id}} |\rho_1 |\psi_{\mathrm{id}} \rangle}$,
and the probability to be in $\mathcal{X}_1$, 
${P_{\mathcal{X}_1}=\mathrm{Tr}(\rho_1)}$.
From Eqs.~(\ref{eq-error1}) and (\ref{eq-error2}), 
we write their evolutions per Clifford gate as
\begin{align}
P_{\mathcal{X}_1} \! (m+1)
=&
\left( 1-L_1-L_2 \right) \! 
P_{\mathcal{X}_1} \! (m) + L_2,
\label{eq-P1}
\\
P_{\mathrm{id\,}} \! (m+1)
=&
\left( 1-L_1 \right) \! \left( 1-p_D \right) \! P_{\mathrm{id\,}} \! (m) 
\nonumber \\
&+ 
\frac{\left[ (1-L_1) p_D-L_2 \right] \! 
P_{\mathcal{X}_1} \! (m) + L_2}{d}.
\label{eq-P2}
\end{align}
Using these relations recursively, 
we obtain the survival probabilities after $m$ Clifford gates followed by 
its ideal inversion as follows:
\begin{align}
&
P_{\mathcal{X}_1} \! (m)
=
A+B \lambda_L^m,
\label{eq-AB}
\\
&
P_{\mathrm{id\,}} \! (m) - \frac{P_{\mathcal{X}_1} \! (m)}{d}
=
C \lambda_r^{m}, 
\label{eq-C}
\end{align}
where 
\begin{align}
\lambda_L&=1-L_1-L_2,
\\
A &= \frac{L_2}{L_1+L_2},
\\
B &= P_{\mathcal{X}_1} \! (0)-A 
\\
\lambda_r&=
\left( 1-L_1 \right) \! \left( 1-p_D \right),
\\
C &= P_{\mathrm{id\,}} \! (0) - \frac{P_{\mathcal{X}_1} \! (0)}{d}. 
\end{align}
It is remarkable that the left-hand side of Eq.~(\ref{eq-C}) 
can be expressed by a single exponential decay without a constant term.
Also note that the state preparation errors, $P_{\mathrm{id\,}} \! (0)$ 
and $P_{\mathcal{X}_1} \! (0)$, 
are included only in $B$ and $C$, 
and therefore we can estimate 
the error-model parameters separately from the state preparation errors.

Finally, we consider the errors in the inversion of $m$ Clifford gates and the measurement.
If they were also isotropic and described 
similarly to Eqs.~(\ref{eq-error1}) and~(\ref{eq-error2}),
the results could still be fitted to
Eqs.~(\ref{eq-AB}) and~(\ref{eq-C}) 
following the relations in Eqs.~(\ref{eq-P1}) and~(\ref{eq-P2}). In RB, however, we set the initial and final states to the ground state $|\widetilde{0000}\rangle$, 
leading to biased~(anisotropic) errors.
We model the experimentally measured probabilities, 
$P_{\mathcal{X}_1}^M \! (m)$ and $P_{\mathrm{id\,}}^M \! (m)$, with 
the biased errors as follows:
\begin{align}
P_{\mathcal{X}_1}^M \! (m)
=&
P_{\mathcal{X}_1} \! (m) 
+
L_{20} \! \left[ 1-P_{\mathcal{X}_1} \! (m) \right],
\label{eq-PSRB1}
\\
P_{\mathrm{id\,}}^M \! (m)
=&
P_{\mathrm{id\,}} \! (m)
+
L_{20} \!\left[ 1-P_{\mathcal{X}_1} \! (m) \right]
\nonumber \\
&+
\gamma \! \left[ P_{\mathcal{X}_1} \! (m) -  P_{\mathrm{id\,}} \! (m) \right] \! ,
\label{eq-PSRB2}
\end{align}
where $L_{20}$ and $\gamma$ are the rates of the population transfers 
from the leakage subspace $\mathcal{X}_2$ and 
the subspace of $\mathcal{X}_1$ orthogonal to $|\widetilde{0000}\rangle$, respectively, to $|\widetilde{0000}\rangle$.
Note that isotropic errors have already been included in 
$P_{\mathcal{X}_1} \! (m) $ and $P_{\mathrm{id\,}} \! (m)$.

We thus obtain
\begin{align}
&
P_{\mathcal{X}_1}^M \! (m)
=
A_M + B_M \lambda_L^m,
\label{eq-ABSRB}
\\
&
P_{\mathrm{id\,}}^M \! (m) 
- \frac{P_{\mathcal{X}_1}^M \! (m)}{d}
=
C_M \lambda_r^{m} + D_M, 
\label{eq-CDSRB}
\end{align}
where 
\begin{align}
A_M &= \left( 1-L_{20} \right) \! A + L_{20} 
\simeq \frac{L_2}{L_1+L_2},
\\
B_M &= \left( 1-L_{20} \right) \! B,
\\
C_M &= \left( 1-\gamma \right) \! C,
\\
D_M &= \frac{d-1}{d} \! \left[ 
\left( \gamma - L_{20} \right) \! P_{\mathcal{X}_1} \! (m) + L_{20} \right]
\nonumber \\
&\simeq
\frac{d-1}{d} \! \left[ 
\left( \gamma - L_{20} \right) \! \bar{P}_{\mathcal{X}_1} + L_{20} \right] \! ,
\end{align}
where $\bar{P}_{\mathcal{X}_1}$ is the average value of 
$P_{\mathcal{X}_1} \! (m)$ with respect to $m$.
This approximation, which allows us to regard $D_M$ as a constant and consequently avoid unreliable fitting with double exponential decays, 
is valid because the dropped term $[(d-1)/d](\gamma - L_{20})
[P_{\mathcal{X}_1} \! (m) - \bar{P}_{\mathcal{X}_1}] $ is small 
compared to the main term $C_M \lambda_r^{m}$ 
for the estimation of $\lambda_r$ with Eq.~(\ref{eq-CDSRB}). 
Note that ${|\gamma - L_{20}| \ll 1}$ and $|P_{\mathcal{X}_1} \! (m) - \bar{P}_{\mathcal{X}_1}|
\simeq |P_{\mathcal{X}_1}^M \! (m) - \bar{P}_{\mathcal{X}_1}^M|$ 
is about one order of magnitude smaller than $\bar{P}_{\mathcal{X}_1} \simeq \bar{P}_{\mathcal{X}_1}^M$ [see Fig.~\ref{Fig4:wide} (b)].

The LRB protocol used in this work is as follows.
In the SRB part, 
we determine 
$\lambda_L^{\mathrm{SRB}}$, $\lambda_r^{\mathrm{SRB}}$, 
$A_M^{\mathrm{SRB}}$, $B_M^{\mathrm{SRB}}$, $C_M^{\mathrm{SRB}}$, 
and $D_M^{\mathrm{SRB}}$
by fitting to the experimental data of 
$P_{\mathcal{X}_1}^{M, \mathrm{SRB}} \! (m)$
and $P_{\mathrm{id\,}}^{M, \mathrm{SRB}} \! (m) - 
P_{\mathcal{X}_1}^{M, \mathrm{SRB}} \! (m)/d$ 
according to Eqs.~(\ref{eq-ABSRB}) and~(\ref{eq-CDSRB}).
The leakage and seepage rates are obtained as 
\begin{align}
L_1^{\mathrm{SRB}}
&=
\left( 1- \lambda_L^{\mathrm{SRB}} \right) \! 
\left( 1- A_M^{\mathrm{SRB}} \right),
\\
L_2^{\mathrm{SRB}}
&=
\left( 1- \lambda_L^{\mathrm{SRB}} \right) \! A_M^{\mathrm{SRB}}.
\end{align}
In the interleaved RB~(IRB) part, 
we similarly obtain 
$\lambda_L^{\mathrm{IRB}}$, 
$\lambda_r^{\mathrm{IRB}}$, 
 $L_1^{\mathrm{IRB}}$, and $L_2^{\mathrm{IRB}}$ by fitting to the experimental data of 
$P_{\mathcal{X}_1}^{M, \mathrm{IRB}} \! (m)$
and $P_{\mathrm{id\,}}^{M, \mathrm{IRB}} \! (m) - 
P_{\mathcal{X}_1}^{M, \mathrm{IRB}} \! (m)/d$.
Assuming the CZ-gate error model similar to Eqs.~(\ref{eq-error1}) and (\ref{eq-error2}), 
parameters for the CZ gate are determined as
\begin{align}
&
1-L_1^{\mathrm{CZ}}
=
\frac{1-L_1^{\mathrm{IRB}}}{1-L_1^{\mathrm{SRB}}},
\label{eq-CZ1}\\
&
\lambda_r^{\mathrm{CZ}}
=
\left( 1-L_1^{\mathrm{CZ}} \right) \!
\left( 1-p_D^{\mathrm{CZ}} \right) \!
=
\frac{\lambda_r^{\mathrm{IRB}}}{\lambda_r^{\mathrm{SRB}}}.
\label{eq-CZ2}
\end{align}
Finally, the average fidelity of the CZ gate is give by
\begin{align}
\bar{F}
&=
\left( 1-L_1^{\mathrm{CZ}} \right) \! 
\left( 1-p_D^{\mathrm{CZ}} + \frac{p_D^{\mathrm{CZ}}}{d} \right)
\nonumber \\
&=
\frac{d-1}{d} \lambda_r^{\mathrm{CZ}} + \frac{1-L_1^{\mathrm{CZ}}}{d},
\end{align}
which can be calculated with the experimentally determined values 
from Eqs.~(\ref{eq-CZ1}) and (\ref{eq-CZ2}).

The average fidelity is approximately expressed as
\begin{align}
\bar{F}
\simeq
1-\frac{d-1}{d} p_D^{\mathrm{CZ}} - L_1^{\mathrm{CZ}}.
\end{align}
That is, the infidelity is given by 
the sum of the depolarizing-induced error 
$r_D^\mathrm{CZ} \equiv (d-1) p_D^\mathrm{CZ}/d$ in the computational subspace 
and the leakage error $L_1^\mathrm{CZ}$.
It is also notable that the average fidelity can be expressed as 
\begin{align}
\bar{F}
=
1-\frac{d-1}{d} \! \left( 1- \lambda_r^{\mathrm{CZ}} \right) \! - \frac{L_1^{\mathrm{CZ}}}{d}.
\end{align}
Since the decay rate $\lambda_r^{\mathrm{CZ}}$ roughly corresponds to 
that estimated in the widely used RB 
neglecting leakage errors, 
the LRB gives the fidelity lower by ${L_1^{\mathrm{CZ}}/d}$ 
than such RB. 

\section{Incoherent error}\label{incoherent error}

While the coherent error of a qubit gate can be minimized through the optimization of control pulses, incoherent errors stemming from various noise sources pose a challenge as they are difficult to mitigate during measurement procedures. The decoherence of a qubit is typically intricate, such as qubit relaxation induced by charge noise~\cite{yan2016flux}, quasiparticles~\cite{PhysRevLett.106.077002}, and two-level systems~\cite{muller2019towards}. Identifying these distinct noise sources is crucial for enhancing qubit coherence and reducing incoherent errors. However, delving into this complexity is beyond the scope of this work. Instead, we partition the impact of distinct decoherence channels, consisting of relaxation and pure dephasing, to facilitate our error analysis.

We start from the 1/\textit{f} flux noise, which is easily estimated experimentally~\cite{PhysRevApplied.13.054079}. Furthermore, its induced incoherent error can be simulated using a Monte Carlo method~\cite{PhysRevX.11.021058}. The error induced by flux noise can be deemed negligible, as demonstrated below, which simplifies the subsequent analysis. We note that various sources of 1/\textit{f}-type noise, including charge noise, contribute to qubit dephasing. However, in a tunable scheme involving transmons, the dominant source is reasonably considered to be flux noise, induced by the fluctuation of the magnetic field in the DTC's loop. Subsequently, an error channel $\mathcal{E}$ described by Kraus operators is employed to estimate the incoherent error induced by qubit energy relaxation and pure dephasing. Given a quantum error channel $\mathcal{E}$ described by Kraus operators as 
($I$ denotes the identity operator and $\rho$ denotes the state density)
\begin{equation}
    \mathcal{E}(\rho) = \sum_k K_k \rho K_k^{\dagger},~
\sum_k K_k^{\dagger} K_k = I,
\end{equation}
the average fidelity for $d$-dimensional systems is given 
by~\cite{PhysRevX.13.031035, pedersen2007fidelity}
\begin{equation}
F_{\mathrm{ave}}
=\frac{1}{d(d+1)}
\sum_k
\left[
\mathrm{Tr} \! \left( K_k^{\dagger} K_k \right) 
+ 
 \left| \mathrm{Tr} \! \left( K_k \right) \right|^2 
\right],
\label{eq-formula}
\end{equation}
where $d$ is 2 and 4, respectively, for single-qubit and two-qubit systems.

\subsection{Flux-noise-induced error}\label{flux noise induced error}

\begin{figure}
\includegraphics[scale=1.0]{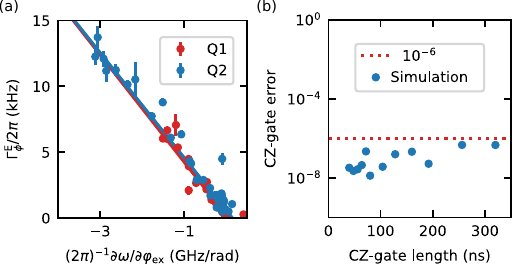}
\caption{\label{FigS10:wide} (a) Dephasing rate calculated from Hahn-echo measurements versus the flux sensitivity of the qubit frequency. Error bars are calculated from the fitting errors. The linear fits depicted by the red and blue solid lines are for Q1 and Q2, respectively. (b) Simulated CZ-gate error versus the CZ-gate length due to the 1/\textit{f} flux noise.
}
\end{figure}

We reused the data obtained in Appendix~\ref{Qubit coherence} for the flux noise calculation. The measurement data for $T_2^\mathrm{E}$ is reanalyzed using the formula: $P_{|1\rangle} = A + Be^{-\Gamma_{\text{exp}}t - (\Gamma_\Phi^\mathrm{E} t)^2}$, where $\Gamma_{\text{exp}}$ mainly accounts for the effects from $T_1$ and white noise, while $\Gamma_\Phi^\mathrm{E}$ characterizes the impact of 1/\textit{f} flux noise. With the frequency sensitivity of a qubit to the flux in the DTC loop, the decay rate induced by the 1/\textit{f} flux noise is determined as~\cite{PhysRevApplied.13.054079}
\begin{equation}
    \Gamma_\Phi^\mathrm{E}=\left( \frac{2\pi}{\Phi_0} \right) \! \sqrt{A_\Phi \ln 2}\left|\frac{\partial \omega}{\partial \varphi_\mathrm{ex}}\right|,
\end{equation}
where $\sqrt{A_\Phi}$ is the amplitude of the flux noise spectra $S_\Phi = A_\Phi / |\omega|$. By fitting $\Gamma_\Phi^\mathrm{E}$ against its frequency sensitivity~($\partial \omega /\partial \varphi_\mathrm{ex}$) for both Q1 and Q2 [Fig.~\ref{FigS10:wide}(a)], the extracted flux noise amplitudes are found to be $\sqrt{A_\Phi} \sim 4.89$ \textmu$\Phi_0$ and $\sim 4.79$~\textmu$\Phi_0$, respectively. It is reasonable for the two qubits to exhibit similar flux noise, as their respective measured flux noise primarily originates from the shared DTC's loop.

Given the average 1/\textit{f} flux-noise amplitude $\sqrt{A_\phi} \sim$~4.84~\textmu$\Phi_0$, a Monte Carlo simulation can be conducted to capture the induced incoherent noise in the CZ gate implementation~\cite{PhysRevX.11.021058}. We initially designed an optimized pulse shape for the CZ gate theoretically for the simulation. Subsequently, we conducted 1000 simulations, where in each iteration, an offset flux sampled from the measured flux noise spectra was added to the optimized pulse. The induced error was calculated as $(1-F_{\mathrm{sim}}^{\mathrm{noise}}) - (1-F_{\mathrm{sim}}^{\mathrm{ideal}})$, where the $F_{\text{sim}}^{\mathrm{ideal/noise}}$ is the simulated gate fidelity~\cite{PhysRevApplied.18.034038} with/without sampled noise. The simulated CZ-gate errors contributed by the 1/\textit{f} flux noise are obtained $<$$10^{-6}$ for the CZ-gate length within 40--320 ns~[Fig.~\ref{FigS10:wide}(b)], which can be considered negligible at the current stage. The simulation result aligns with the findings in Ref.~\citenum{PhysRevX.11.021058}, suggesting that the impact of long-time correlated flux noise is similarly negligible, particularly with the implementation of a short gate length in the DTC scheme.

\subsection{Relaxation-induced error}\label{relaxation induced error}

Denoting the amplitude damping as $p_1 = 1 - e^{-t / T_1} \simeq t / T_1$, with an operation time $t\ll T_1$, the Kraus operators describing the qubit relaxation of a two-qubit system are as follows~\cite{nielsen2010quantum}:

\begin{align}
K_0^{(1)} 
&= 
\begin{pmatrix}
1 & 0 \\ 0 & \sqrt{1-p_1^{(1)}}
\end{pmatrix}
\otimes I \nonumber \\
&=
\begin{pmatrix}
1 & 0 & 0 & 0 \\ 0 & 1 & 0 & 0 \\ 0 & 0 & \sqrt{1-p_1^{(1)}} & 0 \\  0 & 0 & 0 & \sqrt{1-p_1^{(1)}}
\end{pmatrix}, \nonumber\\
\end{align}
\begin{align}
K_1^{(1)} 
&= 
\begin{pmatrix} 
0 & \sqrt{p_1^{(1)}} \\ 0 & 0
\end{pmatrix}
\otimes I  \nonumber \\
&= 
\begin{pmatrix}
0 & 0 & \sqrt{p_1^{(1)}} &  0 \\ 0 & 0 & 0 & \sqrt{p_1^{(1)}} \\ 0 & 0 & 0 & 0 \\  0 & 0 & 0 & 0
\end{pmatrix}, \nonumber\\
\end{align}
\begin{align}
K_0^{(2)} 
&= 
I \otimes 
\begin{pmatrix}
1 & 0 \\ 0 & \sqrt{1-p_1^{(2)}}
\end{pmatrix} \nonumber \\
&= 
\begin{pmatrix}
1 & 0 & 0 & 0 \\ 0 & 1 & 0 & 0 \\ 0 & 0 & \sqrt{1-p_1^{(2)}} & 0 \\  0 & 0 & 0 & \sqrt{1-p_1^{(2)}}
\end{pmatrix}, \nonumber\\
\end{align}
\begin{align}
K_1^{(2)} 
&= 
I \otimes 
\begin{pmatrix}
0 & \sqrt{p_1^{(2)}} \\ 0 & 0
\end{pmatrix} \nonumber \\
&= 
\begin{pmatrix}
0 &  \sqrt{p_1^{(2)}} & 0 & 0 \\ 0 & 0 & 0 & 0 \\ 0 & 0 & 0 & \sqrt{p_1^{(2)}} \\  0 & 0 & 0 & 0
\end{pmatrix},
\end{align}
where $K_k^{(q)}$ and $p_1^{(q)}$ correspond to the $q$th qubit. 

Using the above two-qubit Kraus operators and the formula in Eq.~(\ref{eq-formula}),
we obtain the average incoherent error due to relaxation
of the $q$th qubit in the two-qubit systems: 
\begin{align}
1-F_{\mathrm{ave}}^{(q)}
= 1-
\frac{3-p_1^{(q)}+2\sqrt{1-p_1^{(q)}}}{5}
\simeq \frac{2}{5} \frac{t}{T_1^{(q)}}.
\label{eq-damping2}
\end{align}

\subsection{Pure-dephasing-induced error}\label{Pure dephasing induced error}

Since the flux noise is negligible as simulated above, we focus solely on the predominant contribution of pure dephasing from white noise. This contribution results in an exponential decay of the decoherence signal, rather than a Gaussian profile. Therefore, with $p_\phi=\left(1-e^{-t / T_\phi}\right) / 2 \simeq t /\left(2 T_\phi\right)$ denoting the phase-flip probability, the Kraus operators for pure dephasing of individual qubit in a two-qubit system are as follows~\cite{nielsen2010quantum}:
\begin{align}
K_0^{(1)} 
&= 
\sqrt{1-p_{\phi}^{(1)}}
\begin{pmatrix}
1 & 0 \\ 0 & 1
\end{pmatrix}
\otimes I
=
\sqrt{1-p_{\phi}^{(1)}}
\begin{pmatrix}
1 & 0 & 0 & 0 \\ 0 & 1 & 0 & 0 \\ 0 & 0 & 1 & 0 \\  0 & 0 & 0 & 1
\end{pmatrix}, \nonumber\\
\end{align}
\begin{align}
K_1^{(1)} 
&= 
\sqrt{p_{\phi}^{(1)}} 
\begin{pmatrix}
1 & 0 \\ 0 & -1
\end{pmatrix}
\otimes I
=
\sqrt{p_{\phi}^{(1)}}
\begin{pmatrix}
1 & 0 & 0 & 0 \\ 0 & 1 & 0 & 0 \\ 0 & 0 & -1 & 0 \\  0 & 0 & 0 & -1
\end{pmatrix}, \nonumber\\
\end{align}
\begin{align}
K_0^{(2)} 
&= 
I \otimes 
\sqrt{1-p_{\phi}^{(2)}}
\begin{pmatrix}
1 & 0 \\ 0 & 1
\end{pmatrix}
=
\sqrt{1-p_{\phi}^{(2)}}
\begin{pmatrix}
1 & 0 & 0 & 0 \\ 0 & 1 & 0 & 0 \\ 0 & 0 & 1 & 0 \\  0 & 0 & 0 & 1
\end{pmatrix}, \nonumber\\
\end{align}
\begin{align}
K_1^{(2)} 
&= 
I \otimes 
\sqrt{p_{\phi}^{(2)}} 
\begin{pmatrix}
1 & 0 \\ 0 & -1
\end{pmatrix}
=
\sqrt{p_{\phi}^{(2)}} 
\begin{pmatrix}
1 & 0 & 0 & 0 \\ 0 & -1 & 0 & 0 \\ 0 & 0 & 1 & 0 \\  0 & 0 & 0 & -1
\end{pmatrix},
\end{align}
where $K_k^{(q)}$ and $p_{\phi}^{(q)}$ correspond to the $q$th qubit.

Using the above two-qubit Kraus operators and the formula given in Eq.~(\ref{eq-formula}), we obtain the following formula of the average infidelity for pure dephasing 
of the $q$th qubit in the two-qubit systems, 
\begin{align}
1-F_{\mathrm{ave}}^{(q)}
= 1-
\frac{4p_{\phi}^{(q)}}{5}
\simeq \frac{2}{5} \frac{t}{T_{\phi}^{(q)}}.
\end{align}
Note that this is in the same form as that for the amplitude damping in Eq.~(\ref{eq-damping2}).

In addition to the pure dephasing of individual qubits, we also consider the extra dephasing due to their coupling during the CZ gate, denoted as CZ dephasing. The Kraus operators for the two-qubit CZ dephasing are defined as follows:

\begin{align}
K_0 & =\sqrt{1-p_{\mathrm{CZ}}}\left(\begin{array}{llll}
1 & 0 & 0 & 0 \\
0 & 1 & 0 & 0 \\
0 & 0 & 1 & 0 \\
0 & 0 & 0 & 1
\end{array}\right)=\sqrt{1-p_{\mathrm{CZ}}}\,(I \otimes I), \\
K_1 & =\sqrt{p_{\mathrm{CZ}}}\left(\begin{array}{llll}
1 & 0 & 0 & 0 \\
0 & 1 & 0 & 0 \\
0 & 0 & 1 & 0 \\
0 & 0 & 0 & -1
\end{array}\right) \nonumber\\
& =\sqrt{p_{\mathrm{CZ}}}\, \frac{I \otimes I+Z \otimes I+I \otimes Z-Z \otimes Z}{2},
\end{align}
where $p_{\mathrm{CZ}}=\left(1-e^{-t / T_{\mathrm{CZ}}}\right) / 2 \simeq t /\left(2 T_{\mathrm{CZ}}\right)$. $T_{\mathrm{CZ}}$ is the $\mathrm{CZ}$ dephasing time defined as $\rho_{00,11} \rightarrow e^{-t / T_{\mathrm{CZ}}} \rho_{00,11}, \rho_{10,11} \rightarrow e^{-t / T_{\mathrm{CZ}}} \rho_{10,11}$, and $\rho_{01,11} \rightarrow e^{-t / T_{\mathrm{CZ}}} \rho_{01,11}$.

Using the above two-qubit Kraus operators and the formula given in Eq.~(\ref{eq-formula}), we obtain the average infidelity for the CZ dephasing as follows:
\begin{equation}
    1-F_{\mathrm{ave}}=\frac{3}{5} p_{\mathrm{CZ}} \simeq \frac{3}{10} \frac{t}{T_{\mathrm{CZ}}} .
\end{equation}

\subsection{Summary of incoherent error}\label{summary of incoherent error}

Considering the aforementioned incoherent error analysis, the error induced by flux noise appears negligible given the current CZ-gate error level. Therefore, the incoherent error induced by relaxation and pure dephasing can be modeled as follows:
\begin{equation}
\begin{aligned}
    1-F_{\mathrm{ave}} &\simeq \frac{2}{5} \frac{t}{T_1^{(1)}}+\frac{2}{5} \frac{t}{T_1^{(2)}}+\frac{2}{5} \frac{t}{T_\phi^{(1)}}+\frac{2}{5} \frac{t}{T_\phi^{(2)}}+\frac{3}{10} \frac{t}{T_{\mathrm{CZ}}} \\
    &\equiv \frac{2}{5} \frac{t}{T_{\text{eff}}},
\end{aligned}
\end{equation}
where $T_{\text{eff}}$ denotes the effective coherence time experienced by the two-qubit system.

Despite employing these error models, predicting the exact incoherent error of the CZ gate measured using randomized benchmarking still poses challenges. $T_\mathrm{eff}^\mathrm{est} \approx 67.6\pm 11.4$~\textmu s estimated from $T_1$ and $T_\phi$ of two data qubits measured at the idle bias point is larger than the fitted value $T_\mathrm{eff}^\mathrm{exp} \approx 23.9\pm 1.5$~\textmu s~(Fig.~\ref{Fig5:wide}), which suggests the presence of additional noise induced by the CZ gate. However, accurately extracting the coherence time during the CZ gate proves difficult, given the variability in the noise participation ratio during the gate operation. In particular, a thorough understanding and estimation are necessary to account for the additional contribution of CZ dephasing, which requires further investigation.

In conclusion of this section, we have developed a model to illuminate the potential reasons for the linear relationship between the incoherent gate error and gate length shown in Fig.~\ref{Fig5:wide}. The incomplete explanation of $T_{\text{eff}}$ from the coherence measurement at the idle bias point necessitates further efforts in investigating decoherence during the CZ gate. Furthermore, identifying specific noise sources is crucial for guiding the mitigation of errors induced by these noise channels, which has already become a substantial subject in superconducting qubits and drawn significant attention in recent investigations.

\bibliography{reference}

\end{document}